\documentclass[twocolumn,showpacs,preprintnumbers,amsmath,amssymb,letterpaper]{revtex4}
\usepackage{graphicx}
\usepackage{bm}

\begin{document}
\title{
From high temperature supercondutivity to quantum spin liquid: progress in strong correlation physics
}

\author{Patrick A. Lee}
\address{Department of Physics, Massachusetts Institute of Technology,
Cambridge, Massachusetts 02139}

\begin{abstract}             
This review gives a rather general discussion of high temperature superconductors as an example of a strongly correlated material.  The argument is made that in view of the many examples of unconventional superconductors discovered in the past twenty years, we should no longer be surprised that superconductivity emerges as a highly competitive ground state in systems where Coulomb repulsion plays a dominant role.  The physics of the cuprates is discussed, emphasizing the unusual pseudogap phase in the underdoped region.  It is argued that the resonating valence bond (RVB) picture, as formulated using gauge theory with fermionic and bosonic matter fields, gives an adequate physical understanding, even though many details are beyond the powers of current calculational tools.  The recent discovery of quantum oscillations in a high magnetic field is discussed in this context.  Meanwhile, the problem of the quantum spin liquid (a spin system with antiferromagnetic coupling which refuses to order even at zero temperature) is a somewhat simpler version of the high $T_c$ problem where significant progress has been made recently.  It is understood that the existence of matter fields can lead to de-confinement of the $U(1)$ gauge theory in $2+1$ dimensions, and novel new particles (called fractionalized particles), such as fermionic spinons which carry spin ${1\over 2}$ and no charge, and gapless gauge bosons can emerge to create a new critical state at low energies.  We even have a couple of real materials where such a scenario may be realized experimentally.  The article ends with answers to questions such as: what limits $T_c$ if pairing is driven by an electronic energy scale? why is the high $T_c$ problem hard?  why is there no consensus? and why is the high $T_c$ problem important?

\end{abstract}

\maketitle
\date{\today}

\section{Introduction}
High temperature superconductivity was discovered in cuprates in 1986.\cite{BM8689}
The rapid raising of the transition temperature to well
above the melting point of nitrogen \cite{WAT8708}
shattered the old record of 23~K. Furthermore, the fact that high $T_c$
superconductivity was discovered in a rather unexpected material, a transition
metal oxide, made it clear that some novel mechanism must be at work.  The
intervening years have seen great strides in high  $T_c$ research.  The growth and characterization of cuprate single crystals and thin
films have advanced to the point where sample quality and reproducibility
problems which plagued the field in the early days are no longer issues.  At
the same time, basically all conceivable experimental tools have been applied
to the cuprates.  Indeed, the need for more and more refined data has spurred
the development of experimental techniques such as angle resolved
photoemission spectroscopy (ARPES) and low temperature scanning 
tunneling
microscopy (STM).  Today the cuprate is arguably the best studied material
outside of the semiconductor family and a great deal of facts are known.  It
is also clear that many of the physical properties are unusual, particularly
in the metallic state above the superconductor.  Superconductivity is only one
aspect of a rich phase diagram which must be understood in its totality.

It is often remarked that there is no consensus for the mechanism of high $T_c$ superconductivity.  This may be true but I must emphasize that a lack of consensus is not synonymous with a lack of understanding or lack of progress.  In fact, I will argue that the basic physics of the cuprate family is well understood.   While there are hundreds of high $T_c$ compounds, they all share a layered structure which contains one or more copper-oxygen planes.  The low energy physics of these planes can further be simplified to a model of electrons hopping on a square lattice called the one band Hubbard
model.\cite{A8796,ZR8859} (Many of the details left out in this paper can be found in a more comprehensive review.\cite{LNW0617})
\begin{equation}
H=-\sum_{<i,j>\sigma}t_{ij} c^\dagger_{i\sigma}c_{j\sigma} +
U\sum_{i}n_{i\uparrow}n_{i\downarrow}
\end{equation}
where $c^\dagger_{i\sigma}$ is the creation operator of an electron with spin $\sigma$ on a square lattice,
$n_{i\sigma} = c^\dagger_{i\sigma}c_{i\sigma}$, $t_{ij}$ is the hopping matrix element between sites $i$ and $j$, (we shall denote nearest neighbor hopping by $t$ and further neighbor hopping by $t^\prime$, $t^{\prime\prime}$ etc.), and $U$ is the repulsive energy cost due to screened Coulomb interaction to put two electrons with opposite spin on the same site.  At half-filling (one electron per site), there is a metal-to-insulator transition as the ratio $U/t$ is in increased.  The insulator is called a Mott insulator,\cite{M4916} as opposed to a band insulator, because its existence  is driven by strong repulsion and is not described by band theory.  The latter would require the state to be metallic at half-filling.  The term ``strong correlation physic'' is now used to describe the state of affairs when interaction energy dominates the kinetic energy in controlling the electron motion.  As seen in Fig.~1, for large enough $U/t$ the electrons prefer to be localized on the lattice site because any hopping to reduce the kinetic energy $t$ requires double occupation of some site, which costs $U$.  This insulator is also predicted to be antiferromagnetic (AF), because AF alignment permits virtual hopping to gain an energy $J=4t^2/U$ by second order perturbation theory, whereas hopping is strictly forbidden by Pauli exclusion for parallel spins.  It is now generally agreed that the parent compound of the high $T_c$ cuprate is a Mott insulator.

 \begin{figure}[t]
\centerline{
\includegraphics[width=0.5\textwidth]{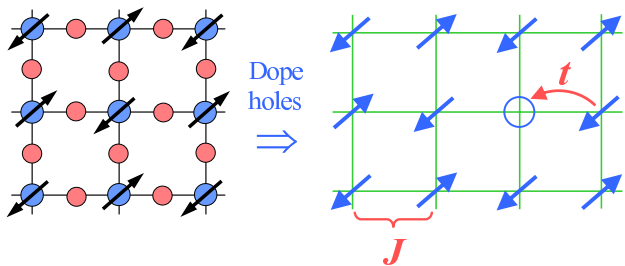}
}
\caption{ 
Structure of the Cu-O layer in high $T_c$ materials.  Copper atoms sit on a square lattice with oxygen atoms in between.  The electronic structure is simplified to a one band model shown on the right, with electrons hopping with matrix element $t$.  There is an antiferromagnetic exchange $J$ between spins on neighboring sites.}
\end{figure}

Things get interesting when electron vacancies (called holes) are introduced into the copper-oxygen layers in a process called hole doping, i.e. a charge reservoir away from the copper-oxygen plane is introduced which removes electrons for the plane.  We denote the concentration of holes by $x$.  The resulting phase diagram is schematically shown in Fig.~2.  The AF order is rapidly destroyed by a few percent of holes, beyond which superconductivity appears.  The transition temperature reaches a maximum around 15\% doping, which is called optimal doping.  The dome-shaped $T_c$ is characteristic of all hole doped cuprates, even though the maximum $T_c$ seems to be clustered around two groups.  It is about 40~K in the La$_{2-x}$Sr$_x$CuO$_4$ (LSCO) family and 93~K and higher in a second family which includes YBa$_2$Cu$_3$O$_{7-\delta}$ (YBCO) and Ba$_2$Sr$_2$CaCu$_2$O$_{9+\delta}$ (Bi-2212).  The highest $T_c$ at ambient pressure of 135~K was reached in HgBa$_2$Ca$_2$Cu$_3$O$_{8+\delta}$ in 1993.\cite{SCG9356}

Before high $T_c$ superconductors were discovered in 1986, almost all superconductors were believed to be $s$-wave BCS superconductors.  The requisite attractive interaction to form pairs is provided by electrons exchanging a phonon.  There were a few potential exceptions in a class of strongly correlated metals called the heavy fermions, but $T_c$ was very low, typically less than 1~K.\cite{
LRS8699}.  This is why the discovery of superconductivity in a system where the repulsion is strong enough to create a Mott insulator was such a surprise.  However, in the intervening years, we know of many examples of non $s$-wave superconductors.  These all occur  in strongly correlated materials and are clearly not driven by electron-phonon coupling.  In the heavy fermion systems, experimental progress has removed any doubt about the non-$s$-wave nature of a large number of compounds.\cite{C0600}  Other new examples include Sr$_2$RuO$_4$ exhibiting triplet pairing which breaks time-reversal symmetry;\cite{MM0357} hydrated cobaltates,\cite{TAK0353}, and several systems near the quantum critical point of a zero temperature ferromagnetic (UGe$_2$)or antiferromagnetic (CePd$_2$Si$_2$, CeIn$_3$)phase transition.\cite{SAX0087,MAT9839}  The transition temperature of these systems remain low, less than 5~K.  Of particular interest to the present discussion is the superconductivity discovered in a series of layered organic molecular solids.\cite{SM0263,B0700}  This is because these compounds also live in the vicinity of the Mott transition.  Unlike the cuprates, they are not doped and remain at half filling.  In these materials the effective hopping parameter $t$ is sensitive to pressure and to the choice of anion molecules.  The ratio $U/t$ can be tuned right through the Mott transition.\cite{KUR0501}  Amazingly it was discovered that when the Mott insulator is destroyed, the system immediately becomes a superconductor, before becoming a metal at even higher pressure.   Furthermore, the transition temperature reaches 11.6~K, the highest known among the organics.  There is also strong evidence that these superconductors have $d$-wave pairing symmetry.\cite{ARA0118}

 \begin{figure}[t]
\centerline{
\includegraphics[width=0.4\textwidth]{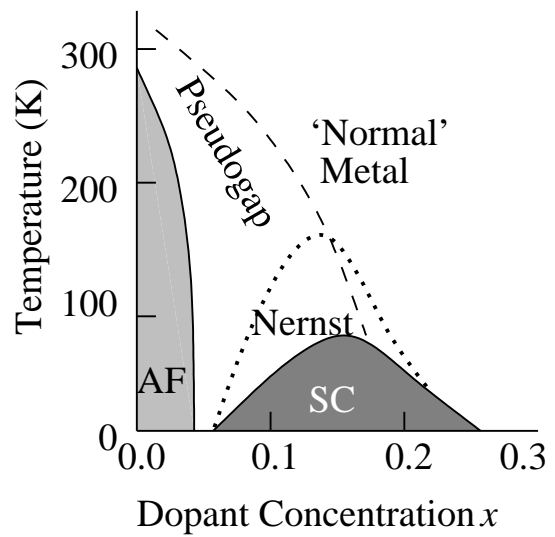}
}
\caption{ 
Schematic phase diagram of high $T_c$ materials.  The antiferromagnet (AF) is rapidly destroyed by doped holes.  The $d$-wave superconductor is subject to strong phase fluctuations below the dotted line, where the proliferation of vortices has been detected by the Nernst effect.  A pseudogap region extends up to high temperatures in the underdoped region.}
\end{figure}

I would argue that 11.6~K for an organic metal qualifies it as an example of a high $T_c$ superconductor!  The reason is that the electronic energy scale for organic solids is much smaller that that for ordinary solids.  For example, the hopping matrix element $t$ is about 0.05~eV compared with 0.4~eV for the cuprates.  Thus the ratio $k_BT_c/t \approx {1\over 40}$ is about the same for both systems.  To emphasize this point, in Fig~3 I have put both materials on the same phase diagram in the parameter space $U/t$ and $x$ of the Hubbard model.  Is the $d$-wave superconductor that appears with doping connected with the one that appears under pressure?  We do not have the answer at present.  My point is that with so many ``unconventional'' examples, our mindset today should be different from that of 20 years ago, and we should be more receptive to the idea that superconductivity may be a highly competitive ground state in a pure repulsive model such as the Hubbard model.

 \begin{figure}[t]
\centerline{
\includegraphics[width=0.45\textwidth]{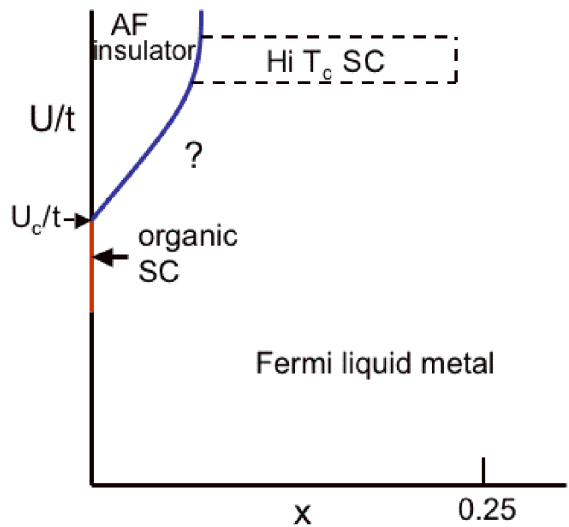}
}
\caption{ 
Location of high $T_c$ cuprates and organic superconductors in the Hubbard model phase diagram.  At half filling, the antiferromagnetic insulator onsets when $U/t$ exceeds a critical value $U_c/t$, where the Mott transition occurs.  High $T_c$ superconductivity occurs when holes are doped into the Mott insulator over a concentration range between 6\% and 25\%.  In certain organic compounds, 12~K superconductivity lives on the boundary between the Mott insulator and the metal.  The ratio $k_BT_c/t$ is about ${1\over 40}$ for both systems.  Whether the two superconducting regions are connected is not known and indicated by the question mark.}
\end{figure}

With that remark  let us return to the cuprates and examine the phase diagram in more detail.  The region between the disappearance of AF and the onset of superconductivity is complicated by disorder effects, even though heroic efforts to make pure samples of YBCO have yielded interesting new information.\cite{DL0601}  We shall not discuss this region further.  The regions of the phase diagram with doping to the left and right of optimal is called underdoped and overdoped, respectively.  The metallic state above $T_c$ in the underdoped region has been under intense study and exhibits many unusual properties not encountered before in any other metal.  This is shown below the dashed line in Fig.~2 and has been called the pseudogap phase.  It is not a well-defined phase in that a definite finite temperature phase boundary has never been found, so the dashed line should be regarded as a cross-over.  There is now broad agreement that the high $T_c$ problem is synonymous with that of doping of a Mott insulator.  It then makes sense to focus on the underdoped region, where the battle line between Mott insulator and superconductivity is drawn.

Since we are interested in the case where $U$ is sufficiently large compared with $t$ for the Hubbard model to be in the Mott insulator phase, it is useful to expand in $t/U$.  The leading order result is the $t$-$J$ model

\begin{equation}
H=P
\left[
\sum_{<ij>,\sigma} t_{ij} c^\dagger_{i\sigma}c_{j\sigma} +
J \sum_{<ij>}
\left(
\bm{S}_i \cdot \bm{S}_j - {1\over 4} n_in_j
\right)
\right]
P .
\end{equation}
The second term is the AF Heisenberg exchange between local spins
$
\bm{S}_i = \frac{1}{2} c^\dagger_{i\alpha}\bm{\sigma}_{\alpha\beta}c_{i\beta}
$
discussed earlier.  The nontrivial part of the $t$-$J$ model resides in the projection operator $P$ which restricts the Hilbert space to exclude the doubly occupied states.  The strong Coulomb repulsion now becomes a constraint of no double occupation.  Compared with the Hubbard model, the Hilbert space is reduced from four states per site to three, namely spin up, spin down or empty.  The parameters of the $t$-$J$ model appropriate for the cuprates is also well established.  $J \sim 0.13$~eV $\sim 1500$~K, $t/J \sim 3$ and $t^\prime/t$ is negative, of order $-0.2$, and is believed to vary somewhat from compound to compound.\cite{PDS0103}

Equations (1) and (2) are deceptively simple looking Hamiltonians which have defied accurate numerical or analytic solution.  Nevertheless, the belief among many workers in the field is that they contain the rich physics of the high $T_c$ phase diagram.  The situation is not unlike QCD, where the Lagrangian is known, but precise understanding of confinement and the mass spectrum has just begun to emerge from quantum Monte Carlo after decades of hard work.  To make matters worse, the high $T_c$ problem at finite doping is analogous to the QCD problem with finite quark density,\cite{MR0600} where accurate numerical solution is so far not possible due to the fermion sign problem.  On the other hand, unlike the quark-gluon problem the high $T_c$ problem has a lot more experimental constraint and input.  As a result we know a lot about the high $T_c$ phenomenology which severely limits the theoretical options.

\section{Simple Physical Picture and the Pseudogap Phenomenology}
Let us start with some simple common sense arguments to gain some insight into the nature of the problem of a doped Mott insulator.  Consider a single hole hoping in an AF background as shown in Fig.~1.  After one hop we find a spin surrounded by ferromagnetic neighbors, costing an energy of ${3\over 2} J$ from the 3 ferromagnetic bonds if the spins are treated as classical $S = {1\over 2}$.  There is a competition between the exchange energy $J$ and the desire of the hole to hop in order to gain the kinetic energy $t$ per hole.  For large enough doping the kinetic energy wins and we expect a metallic state with some short range AF correlation.  By comparing $xt$ and $J$, we expect this to onset at $x \sim {J\over t} \sim {1\over 3}$, in good agreement with the experimental finding.  This state should be a Fermi liquid state.  There is a powerful theorem in Landau Fermi liquid theory commonly called the Luttinger theorem\cite{AGD6500} which states that the area of the Fermi surface is the same as that of free fermions, i.e., it is determined by the total density of electrons in the unit cell. In our case the area is ${1\over 2}(1-x)A_{BZ}$ where $A_{BZ} = (2\pi/a)^2$ is the area of the Brillouin zone (BZ).  This is exactly what is found experimentally.  In Fig.~4(d) we show an example of the measured Fermi surface.  The precise shape can be fitted with a hopping model with further neighbor hopping.

The opposite limit of a few holes $(x \ll 1)$ hopping in an AF background is less trivial, but by now reasonably well understood.  The competition with the AF exchange causes the effective hopping matrix element to be renormalized downward from $t$ to $J$.\cite{KLR8980,SRV8893,LM9225}  The quasiparticle nevertheless manages to form coherent bands.  The bands have minima at $\left( \pm{\pi\over 2a}, \pm{\pi \over 2a}\right)$.\cite{SS8867}  With finite doping the Fermi surfaces are ellipses centered at 
$\left( \pm{\pi\over 2a}, \pm{\pi \over 2a}\right)$ as shown in Fig.~4(a).  Note that the unit cell is doubled because of AF ordering and the BZ is reduced to the diamond in Fig.~4(a).  Applying Luttinger theorem to the doubled unit cell, the total area of the Fermi surface in the reduced BZ is now $(1-x)A_{RBZ}$ where $A_{RBZ} = {1\over 2}A_{BZ}$.  Therefore we conclude that the area of each ellipse (hole pocket) is ${x\over 4}A_{BZ}$.  Physically it makes sense that transport properties are determined only by $x$ carriers occupying small Fermi pockets.  The theory of a few holes in AF background is quite well developed, and recently papers applying effective field theory approach  borrowed from the particle physics literature are particularly notable.\cite{MS0701,BRU0652}

We have good understanding of $x \ll 1$ and $x \gtrsim {1\over 3}$.  What happens in between?  Here we run into a dilemma. We know that AF order is destroyed for $x \gtrsim 0.03$, beyond which points we have no indication of unit cell doubling.  If Fermi liquid theory were to hold, what would happen to Luttinger theorem?   Recall that the nice physical picture of small hole pockets rely on the unit cell doubling.  Once that is absent, Luttinger theorem forces us to have a ``large'' F.S., i.e. one with area proportional to $1-x$.  In that case it will be difficult to see how transport properties will continue to look as if it is given by $x$ holes.  We note that while the original derivation of the Luttinger theorem was perturbative in the interaction strength, the modern derivation by Oshikawa\cite{O0070,PV0418} is a topological one and relies on very few assumptions, not much beyond the statement that well defined quasiparticles exist.  In principle, the Fermi liquid can develop a heavy mass $\approx {1\over x}$ so that the conductivity spectral weight $n/m^\ast \approx x$, but experimentally there is no evidence of such heavy mass formation.  Parenthetically we point out that the three dimensional example of doped Mott insulator La$_{2-x}$Sr$_x$TiO$_3$ appears to take the heavy mass route.\cite{TOK9326}

It turns out that Nature solves this problem in an extremely clever and unexpected way.  As far as the ground state is concerned, the question is moot because it appears that once AF is destroyed the system becomes superconducting and the Luttinger theorem cannot be applied.  What about the normal state above the superconducting $T_c$?  The extensive work using angle resolved photoemission spectroscopy (ARPES) has shown that the gapless excitations lie on an arc.\cite{CNR03,DHS0373}  Anywhere apart from the arc, the excitations are gapped.

 \begin{figure}[t]
\centerline{
\includegraphics[width=0.5\textwidth]{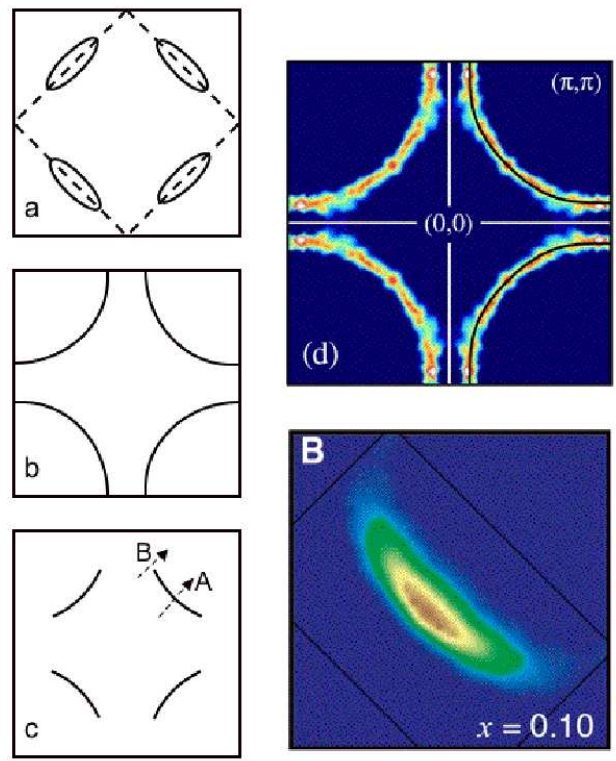}
}
\caption{ 
(a)~Fermi pockets in a doped AF.  Dashed lines indicate the reduced Brillouin zone due to the unit cell doubling of the AF. (b)~Fermi surface of a tight binding model with first and second nearest neighbor hopping. (c)~Schematic picture of the Fermi arcs. the excitations are gapless when path A crosses the arc but are gaped everywhere along path B. (d)~Experimental data showing the Fermi surface in overdoped Tl-2201 ($x=0.25$).  Colors indicate the intensity of low energy excitations.  Data from Plat\'{e} {\em et al.}\cite{PLA0501}  (e)~Experimental data showing the Fermi arc in one quadrant of Figure 4(c) in underdoped Na$_{2-x}$Ca$_x$Cu$_2$O$_2$Cl$_2$ $(x=0.1)$.  Data from K. Shen {\em et al.}\cite{SHE0501}}
\end{figure}

This situation is sufficiently strange that requires a bit more explanation in terms of the experimental observation.  ARPES measures the spectrum of occupied electron states which can be removed by excitations with a photon, i.e. it measures the hole spectral function.  A spectrum is measured at every $\bm{k}$ point.  In a Fermi liquid the spectrum consists of a quasiparticle peak at energy $\varepsilon_{\bm{k}}$.  As one moves along line A in Fig.~4(c), the peak approaches the Fermi energy and disappears as $\varepsilon_{\bm{k}}$ crosses the Fermi surface, thus locating its position.  This is how the Fermi surface in Fig.4(b) is mapped out.  In the underdoped case, what happens is that along line B, the quasiparticle peak (now quite broad) approaches the Fermi energy but stops before crossing it.  Instead it loses weight and disappears.  You might say that this also happens in Fig.4~(a), if path B misses the hole pocket.  The important difference is that in the case of the hole pocket where there is unit cell doubling, in the extended zone scheme we expect the occupied band to live also outside the reduced BZ.    Thus if we follow path A we should see a quasiparticle peak appearing at the second crossing of the ellipse.  Further along path A this peak will then move down in energy away from the Fermi energy.  In other words, the back side of the hole pocket should be visible in ARPES as an occupied quasiparticle state rises up to meet the FS.  The surprise is that in Fig.~4(c), there is no back side to the ellipse, and one is left with what is called a Fermi arc.  The spectrum is gapped everywhere except on the arc, and the gap reaches a maximum near $(\pi,0)$.  In the superconducting state the arc also becomes gapped, leaving a single gapless point along $(\pi,\pi)$ called the nodal point.  In this way the quasiparticle spectrum of a $d$ wave superconductor is smoothly formed out of the pseudogap normal state.  The size of the maximum gap has been measured as a function of doping, and is found to increase with decreasing $x$, in a way which tracks the onset of the pseudogap phase shown in Fig.~2.  (A word of caution: the electron spectral function near $(\pi,0)$, the so called anti-nodal direction, is extremely broad and, unlike the nodal directions, does not show a quasiparticle peak even in the superconducting state in strongly underdoped samples.  So the gap is measured as the pull-back of the leading edge of the spectrum for the Fermi energy.  This picture is corroborated by extensive scanning tunneling microscopy (STM) work, which has excellent energy and spatial resolution, but no momentum space information.   

The Fermi arc and anti-nodal gap is one part of a rich phenomenology associated with the pseudogap phase.  The first evidence of an energy gap comes from NMR data, which found that in underdoped samples the Knight shift, which measures the spin susceptibility, does not follow Pauli's prediction of being temperature independent, but starts dropping around room temperature and has lost about 80\% of its value by the time $T_c$ is reached.\cite{CJS9777}  The gap also shows up in $c$-axis frequency dependent conductivity,\cite{HTL9310} where electrons move from one layer to the next, but the transport within the plane remains metallic, being dominated by the quasiparticles near the Fermi arc.

What is the origin of the large gap at $(\pi,0)$?  One suggestion is that it is simply a $d$-wave superconducting gap and the pseudogap phase should be understood as a superconductor destroyed by strong phase fluctuations.  Phase fluctuation is controlled by the superfluid stiffness which is given by the superfluid density.  Thus $T_c$ should be proportional to the superfluid density, i.e. decreases with decreasing $x$.\cite{EK9534} This picture is made more quantitatively accurate if the reduction of the superfluid density due to thermal excitations of nodal quasiparticles is taken into account.\cite{LNW0617,LW9711}  The opposite trend of the energy gap and $T_c$ is explained, but the origin of such a large superconducting gap close to an insulator becomes even a greater puzzle.  Experimentally this is now ample evidence that the picture of fluctuating superconductivity indeed survives up to much higher temperature compared with a conventional metal: perhaps 2 or 3 times $T_c$.  However, that scale decreases with decreasing doping and does not reach up to the pseudogap temperature.  The key experiment that mapped out this region is the Nernst effect, which is sensitive to mobile superconducting vortices.\cite{WLO0610}  For this reason we refer to this region as the Nernst region in Fig.~2.

If the pseudogap is not a pairing gap, it presents a great challenge for theory, because while the Fermi arc scenario interpolates between the small hole pocket and large Fermi surface beautifully, it is not allowed by conventional band theory or Fermi liquid theory.  Fermi surfaces do not simply terminate.  This is a situation not encountered before in solid state physics.  This is why the psuedogap phenomenon is considered one of the central mysteries of the high $T_c$ story. We shall return to this issue in section V.

\section{The RVB Picture}

While the pseudgap phenomena is really strange, to a large extent it was anticipated by theory.  Here I am referring to the concept of the resonating valence bond (RVB) introduced by P.W. Anderson\cite{A8796} and the slave boson mean field theory and elaborations which followed.  Here we provide a brief review.

We explained in the last section that the N\'{e}el spin order is incompatible
with hole hopping.  The question is whether there is another arrangement of
the spin which achieves a better compromise between exchange energy and the
kinetic energy of the hole.  For $S={1\over 2}$ it appears possible to take
advantage of the special stability of the singlet state. The ground state of
two spins $S$ coupled with antiferromagnetic Heisenberg exchange is a spin
singlet with energy $-S(S+1)J$.  Compared with the classical large  spin
limit, we see that quantum mechanics provides an additional stability in the
term unity in $(S+1)$ and this contribution is strongest for $S={1\over 2}$.
Let us consider a one-dimensional spin chain.  A N\'{e}el ground state with
$S_z = \pm {1\over 2}$ gives an energy of $-{1\over 4}J$ per site.  On the
other hand, a simple trial wavefunction of singlet dimers already gives a
lower energy of $-{3\over 8}J$ per site.  This trial wavefunction breaks
translational symmetry and the exact ground state can be considered to be a
linear superposition of singlet pairs which are not limited to nearest
neighbors, resulting in a ground state energy of 0.443~J.  In a square and
cubic lattice the N\'{e}el energy is $-{1\over 2}J$ and $-\frac34J$ per site,
respectively, while the dimer variational energy stays at $-{3\over 8}J$.  It
is clear that in a 3D cubic lattice, the N\'{e}el state is a far superior
starting  point, and in two dimensions the singlet state may present a serious
competition.  Historically, the notion of a linear superposition of spin
singlet pairs spanning different ranges, called the resonating valence bond
(RVB), was introduced by 
\cite{A735} and 
\cite{FA7432} as a
possible ground state for the $S={1\over 2}$ antiferromagnetic Heisenberg
model on a triangular lattice.  The triangular lattice is of special interest
because an Ising-like ordering of the spins is frustrated.  Subsequently, it
was decided that the ground state forms a $\sqrt{3} \times \sqrt{3}$
superlattice where the moments lie on the same plane and form $120^\circ$
angles between neighboring sites.\cite{HE8831}  

Soon after the discovery of high $T_c$ superconductors, Anderson
\cite{A8796}
revived the RVB idea and proposed that with the introduction of holes the
N\'{e}el state is destroyed and the spins form a superposition of singlets.
The vacancy can hop in the background of what he envisioned as a liquid of
singlets and a better compromise between the hole kinetic energy and the spin
exchange energy may be achieved.  Many elaborations of this idea followed,
but here we argue that the basic physical picture described above gives a
simple account of the pseudogap phenomenon.  The singlet formation explains
the decrease of the uniform spin susceptibility. The vacancies are responsible for transport in the
plane.  The conductivity spectral weight in the $ab$ plane is given by the
hole concentration $x$ and is unaffected by the singlet formation.  On the
other hand, for $c$-axis conductivity, an electron is transported between
planes.  Since an electron  carries spin ${1\over 2}$, it is necessary to
break a singlet.  This explains the gap formation in $\sigma_c(\omega)$ and
the energy scale of this gap should be correlated with that of the uniform
susceptibility.  In photoemission, an electron leaves the solid and 
reaches
the detector, the pull back of the leading edge simply reflects the energy
cost to break a singlet.  

A second concept associated with the RVB idea is the notion of spinons and
holons, and spin charge separations.  Anderson  postulated that the spin
excitations in an RVB state are $S={1\over 2}$ fermions which he called
spinons.  This is in contrast with excitations in a N\'{e}el state which are
$S = 1$ magnons or $S = 0$ gapped singlet excitations.
 \begin{figure}[t]
\centerline{
\includegraphics[width=0.5\textwidth]{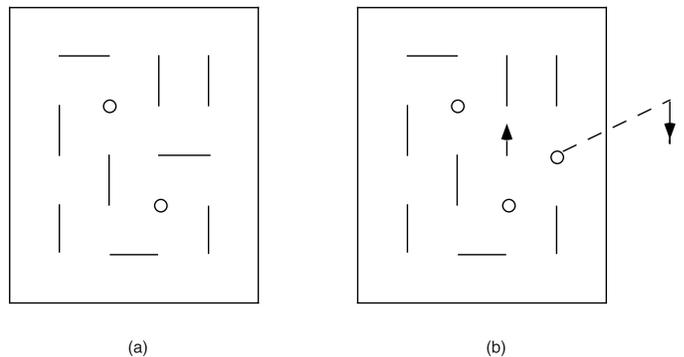}
}
\caption{
A cartoon representation of the RVB liquid or singlets.  Solid bond represents a spin singlet
configuration and circle represents a vacancy.  In (b) an electron is removed from the plane in
photoemission or $c$-axis conductivity experiment.  This necessitates the breaking of a singlet. }
 \label{RVB}
\end{figure}

Initially the spinons are  suggested to form a Fermi surface, with Fermi
volume equal to that of $1-x$ fermions.\cite{BZA8773}  Later it was proposed that the Fermi
surface is gapped to form $d$-wave type structure, with maximum gap near
$(0,\pi)$.\cite{KL8842}  This $\bm{k}$ dependence of the energy gap is needed to explain the
momentum dependence observed in photoemission.

The concept of spinons is a familiar one in one-dimensional spin chains where
they are well understood to be domain walls.  In two dimensions the concept is
a novel one  which does not involve domain walls.  Instead, a rough physical
picture is as follows.  If we assume a background of short range singlet
bonds, forming the so-called short-range RVB state, a cartoon of the spinon is
shown in Fig. \ref{RVB}.  If the singlet bonds are ``liquid,'' two $S={1\over
2}$ formed by breaking a single bond can drift apart, with the liquid of
singlet bonds filling in the space between them.  They behave as free
particles and are called spinons.  The concept of holons follows naturally
\cite{KRS8765}
as the vacancy left over by removing a
spinon.  A holon carries charge $e$ but no spin.

\section{Projected Wavefunction, Slave Boson and the Gauge Theory Formulation of the RVB Picture}

Is there any calculated tool or mathematical formalism to put some meat into the physical picture of RVB described in the last section?  As far as computation is concerned, the use of the projected wavefunction has enjoyed considerable success.  The idea is to write down a trial wavefunction of the type

\begin{equation}
\Psi = P_G \phi
\end{equation}
where $P_G = \prod_i(1-n_{i\uparrow} n_{i\downarrow})$
is called the Gutzwiller projection and $\phi$ is any Hartree Fock or BCS wavefunction, usually suggested by mean field theory described below.  The role of the Gutzwiller projection is to remove all doubly occupied states in $\phi$.  Equation~(3) is a suitable variational wavefunction for the $t$-$J$ model because the constraint is satisfied by definition, and its expectation values and correlation functions can be computed by efficient Monte Carlo algorithms.\cite{G8953,Edegger}  The mean field parameters can be treated as variational parameters.  The projection wavefunction gives excellent ground state energy and sublattice magnetization at half-filling, capturing the important quantum fluctuations of the N\'{e}el ordered state.  With doping it predicted correctly the $d$-wave pairing ground state,\cite{G8831} even though the prediction of the co-existence of superconductivity with AF up to $x\approx 0.11$ is not in agreement with experiment.  Putting aside the question of whether it is the ground state, a comparison of the physical properties of the projected $d$-wave pairing states with a variety of experiments was successfully made.\cite{PRT0404}  The trial wavefunction can be further improved by the Lanczos method of repeatedly hitting it with the Hamiltonian.  There is some controversy as to whether the ground state of the $t$-$J$ model with nearest-neighbor hopping only is a $d$-wave superconductor,\cite{SMB0202,SCL9894} but it is clear that superconductivity is a highly competitive state, as found by other numerical methods such as density matrix renormalization group,\cite{WS9953} cluster dynamical mean field theory \cite{Maier} and variational cluster 
approximation.\cite{Senechal,Tremblay}  Recently it was found that introduction of $t^\prime$ considerably stabilizes the $d$-wave superconducting state.\cite{SLE0402}  At present, I would stay that there is strong numerical evidence that the $d$-wave superconductor is a strong contender for the ground state of the $t$-$J$ model.

What about analytic theory and where does the mean field $\phi$ come from?  A useful method is called the Gutzwiller approxmation, which imposes the constraint approximately by treating the available configuration for hopping and exchange in a statistical basis.\cite{ZGR8836}  This is clearly related to the slave-boson method which we discus below. The slave-boson method was developed for the Kondo problem\cite{B7675,C8435}  It has enjoyed great success as the best way to undersand the properties of a remarkable class of materials called the heavy fermion compounds, where Fermi liquid theory has been stretched to the extreme, with effective mass as large as several thousand times the free electron mass.\cite{H9300}  The idea is to write the electron operator as a product of the boson and fermion which carries the spin index

\begin{equation}
c_{\bm i\sigma}^\dagger = f_{\bm i\sigma}^\dagger b_{\bm i}
\end{equation}
with the condition
\begin{equation}
f_{\bm i\uparrow}^\dagger f_{\bm i\uparrow}
+ f_{\bm i\downarrow}^\dagger f_{\bm i\downarrow} + b^\dagger_{\bm i} b_{\bm i} = 1.
\end{equation}
This constraint can be enforced with a Lagrangian multiplier $\lambda_{\bm i}$.
Note that Eq.~(4) is not an operator
identity and the right-hand-side does not satisfy the fermion commutation relation.
Rather, the requirement is that both sides have the correct matrix elements in
the reduced Hilbert space with no doubly occupied states.  For example, the
Heisenberg exchange term is written in terms of $f^\dagger_{\bm i\sigma}$,
$f_{\bm i\sigma}$ only \cite{BZA8773}
\begin{eqnarray}
{\bm{S}}_{\bm i}\cdot {\bm{S}}_{\bm j} &=& -{1\over 4} f_{\bm i\sigma}^\dagger f_{\bm j\sigma}
f_{\bm j\beta}^\dagger f_{\bm i\beta}  \nonumber   \\
&-& {1\over 4} \left(
f_{\bm i\uparrow}^\dagger f_{\bm j\downarrow}^\dagger - f_{\bm i\downarrow}^\dagger f_{\bm j\uparrow}^\dagger
\right) \left(
f_{\bm j\downarrow} f_{\bm i\uparrow} - f_{\bm j\uparrow} f_{\bm i\downarrow}
\right)  \nonumber \\ 
&+& {1\over 4} \left( f_{\bm i\alpha}^\dagger f_{\bm i\alpha}  \right) .
\end{eqnarray}
We then decouple the exchange term in both the
particle-hole and particle-particle channels via the Hubbard-Stratonovich (HS) transformation.  

Then the partition function is written in the 
form
\begin{equation}
Z = \int{D f D f^\dagger Db D\lambda D\chi D\Delta} \exp \left(
-\int^\beta _0 d\tau L_1
\right)
\end{equation}
where
\begin{eqnarray}
L_1 &=& \tilde{J} \sum_{\langle\bm i\bm j\rangle} \left(
|\chi_{\bm i\bm j}|^2 + | \Delta_{\bm i\bm j} |^2 \right)
+\sum_{\bm i\sigma}f_{\bm i\sigma}^\dagger (\partial_\tau - i\lambda_{\bm i})
f_{\bm i\sigma} 
\nonumber\\
&-& \tilde{J} 
\left[
 \sum_{\langle\bm i\bm j \rangle} \chi_{\bm i\bm j}^\ast \left( \sum_{\sigma} f_{\bm i\sigma}^\dagger
f_{\bm j\sigma} \right) + c.c. \right]  
\\
&+& \tilde{J} \left[ \sum_{\langle\bm i\bm j\rangle} \Delta_{\bm i\bm j} \left(
f^\dagger_{\bm i\uparrow}f^\dagger_{\bm j\downarrow} - f^\dagger_{\bm i\downarrow} f^\dagger_{\bm j\uparrow} \right) + c.c. \right] \nonumber \\
&+&\sum_{\bm i} b_{\bm i}^\ast(\partial_\tau - i\lambda_{\bm i} + \mu_B ) b_{\bm i} 
- \sum_{\bm i\bm j}{t}_{\bm i\bm j}b_{\bm i}b_{\bm j}^\ast f_{\bm i\sigma}^\dagger f_{\bm j\sigma} ,
\nonumber 
\end{eqnarray}
with $\chi_{\bm i\bm j}$ representing fermion hopping and $\Delta_{\bm i\bm j}$
representing fermion pairing corresponding to the two ways of representing the
exchange interaction in terms of the fermion operators. $\tilde{J} = 3J/8$ is chosen to reproduce the mean field self-consistent
equation which is obtained by the Feynman variational principle.
       
Mean field theory corresponds to the saddle
point solution to the functional integral.  The mean field conditions are
\begin{eqnarray}
\chi_{\bm i\bm j} &= \sum_\sigma \langle
f^\dagger_{\bm i\sigma} f_{\bm j\sigma} \rangle \\
\Delta_{\bm i\bm j} &= \langle  f_{\bm i\uparrow}f_{\bm j\downarrow} - f_{\bm i\downarrow}f_{\bm i\uparrow}  \rangle
\end{eqnarray}

Let us write $\chi_{ij} = |\chi_{ij}|e^{ia_{ij}}$ and ignore the fluctuation of the amplitude.  Furthermore, in the last term in Eq.~(8) we replace  $f_{i\sigma}^\dagger f_{j\sigma}$ by $\chi_{ij}$.  Then the rather complicated Lagrangian of Eq.~(8) has a rather simple interpretation.  It describes fermions and bosons hopping on a lattice with hopping matrix element $\chi_{ij}$.  The phase $a_{ij}$ lives on the links, and plays the role of the spatial components of a lattice gauge field, while the $\lambda_i$ fields introduced to enforce the constraint become the time components.  Note that both fermions and bosons are coupled to the same gauge field.  In addition, the fermions may have singlet pairing amplitude given by $\Delta_{ij}$.  Thus we come to the conclusion that the $t$-$J$ model is equivalent to a lattice gauge theory with non-relativistic fermions and bosons coupled with a {\em compact} $U(1)$ gauge field (compactness simply refers to the fact that the gauge field $a_{ij}$ is a phase defined module $2\pi$).  At this point the gauge field has no dynamics.  It fluctuates freely and is in the infinite coupling limit.  Interesting dynamics emerge upon integrating out some of the matter field but one is left with a lattice gauge theory with strong coupling.  The mapping is basically exact, but the question remains as to how to deal with such a model.

Before discussing the importance of gauge fluctuations, let us examine some examples of mean field solutions.

\begin{enumerate}

\item $d$-wave pairing states.  \\
Here  $\chi_{ij} = \chi$ is constant and $\Delta_{ij} = \Delta$ for $(ij)$ bonds along $x$ and $\Delta_{ij} = -\Delta$ for $(ij)$ along $y$, i.e. it has $d$-wave pairing geometry.  Without pairing, the fermions hop on a tight binding band with dispersion

\begin{equation}
\varepsilon_f(\bm k) = -2 \tilde{J}\chi (\cos k_xa + \cos k_ya) .
\end{equation}
With pairing we have the classic $d$-wave dispersion

\begin{equation}
E(k) = 
\sqrt
{
\left(
\varepsilon_f (\bm k) - \mu_f
\right)^2
+
|\Delta_k|^2
}
\end{equation}
where $\mu_f$ is the fermion chemical potential and $\Delta_k = 2\Delta (\cos k_xa -\cos k_ya)$.  The bosons see the same band  dispersion and condense at the bottom of the band minimum at low temperatures.  In mean field theory the Bose condensation temperature is proportional to the boson density $x$.  Below this temperature we have electron pairing, because the BCS order parameter $\langle c_{k\uparrow}c_{-k\downarrow}\rangle = b_0^2\langle f_{k\uparrow}f_{-k\downarrow}\rangle \neq 0$ where $\langle b \rangle = b_0$.  The mean field phase diagram is shown in Fig.~6 and captures some key features of the high $T_c$ phase diagram shown in Fig.~2.  In particular, the $d$-wave superconducting state appears at intermediate doping and a spin gap state (region II) where a $d$-wave-like gap exists for spin excitations but not charge excitations, anticipates many of the properties of the pseudogap phase.

\begin{figure}[t]
\centerline{
\includegraphics[width=0.4\textwidth]{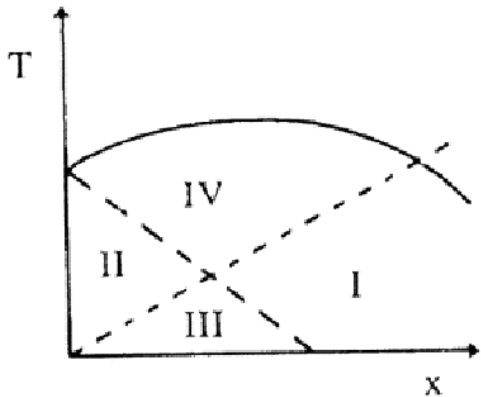}
}
\caption{ 
Schematic phase diagram of the $U(1)$ mean field theory.  The solid line
denotes the onset of the uniform RVB state $(\chi \neq 0)$.  The dashed line
denotes the onset of fermion pairing $(\Delta \neq 0)$ and the dotted line
denotes mean field Bose condensation $(b \neq 0)$.  The four regions are
(I)~Fermi liquid $\chi \neq 0$, $b \neq 0$; (II)~spin gap $\chi \neq 0$,
$\Delta \neq 0$; (III)~$d$-wave superconductor $\chi \neq 0$, $\Delta \neq 0$,
$b \neq 0$; and (IV)~ strange metal $\chi \neq 0$.  From Lee and Nagaosa.\cite{LN9221}}
\end{figure}

\item
The staggered flux state.\\
 Early in the development of the mean field theory, a variety of mean field states were discovered which give identical dispersion. Notable among these is the staggered flux
state 
\cite{AM8874}  In this state the hopping $\chi_{\bm i\bm j}$ is
complex, $\chi_{\bm i\bm j} = \chi_0 \exp \left( i (-1)^{i_x+j_y} \Phi_0 \right) $,
and the phase is arranged in such a way that it describes free fermion hopping
on a lattice with a fictitious flux $\pm 4 \Phi_0$ threading alternative
plaquettes.  Remarkably, the eigenvalues of this problem are identical to that
of the $d$-wave superconductor given by eq.~(12), with $\mu_f = 0$ and 
\begin{equation}
\tan \Phi_0 = {\Delta \over \chi}  .
\end{equation}
The case $\Phi_0 = \pi/4$, called the $\pi$ flux phase, is special in that it
does not break the lattice translation symmetry.  As we can see from
Eq.~(13),
the corresponding $d$-wave problem has a very large energy gap and its
dispersion is shown in Fig.~7.  The key feature is that the energy
gap vanishes at the nodal points located at $\left( \pm{\pi\over 2},
\pm{\pi\over 2} \right)$.  Around the nodal points the dispersion rises
linearly, forming a cone which resembles the massless Dirac spectrum.  For the
$\pi$ flux state the dispersion around the node is isotropic.  For $\Phi_0$
less than $\pi /4$ the gap is smaller and the Dirac cone becomes progressively
anisotropic.    The anisotropy can be characterized by two velocities, $v_F$
in the direction towards $(\pi,\pi)$ and $v_\Delta$ in the direction towards
the maximum gap at $(0,\pi)$.

\begin{figure}[t]
\centerline{
\includegraphics[width=0.4\textwidth]{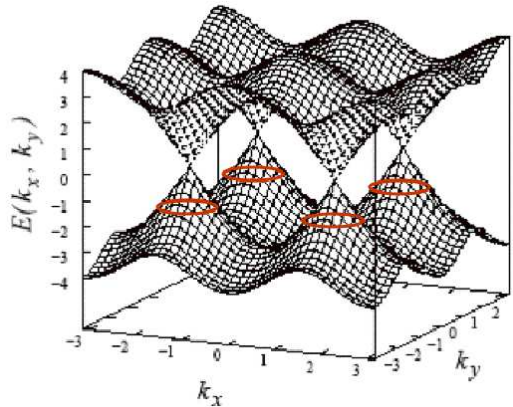}
}
\caption{
The energy dispersion of the staggered flux phase.  Note the massless Dirac spectrum at the nodal points $\left( \pm{\pi\over 2}, \pm{\pi\over 2} \right)$.  Figure shown is for the special case of $\pi$-flux.  In general, the nodal spectra becomes anisotropic.  With doping Fermi pockets are formed when the Fermi energy crosses the energy spectrum.}
\end{figure}

The reason various mean-field theories have the same energy at half-filling was explained by
Affleck {\em et al}.\cite{AZH8845} and Dagotto {\em et al}.\cite{DFM8826}  as being due to a
certain $SU(2)$ symmetry.  It corresponds to the following
particle-hole transformation
\begin{eqnarray}
f_{\bm i\uparrow}^\dagger &\rightarrow& \alpha_{\bm i} f_{\bm i\uparrow}^\dagger + \beta_{\bm i} f_{\bm i\downarrow} \\ \nonumber
f_{\bm i\downarrow} &\rightarrow& -\beta_{\bm i}^\ast f_{\bm i\uparrow}^\dagger + \alpha_{\bm i}^\ast f_{\bm i\downarrow}  .
\end{eqnarray}
Note that the spin quantum number is conserved.  It describes the physical
idea that adding a spin-up fermion or removing a spin-down fermion are the
same state after projection to the subspace of singly occupied fermions.  It
is then not a surprise to learn that the Gutzwiller projection of the $d$-wave
superconductor and that of the staggered flux state gives the same trial
wavefunction, up to a trivial overall phase factor, provided $\mu_f = 0$ and
Eq.~(13) is satisfied.  A simple proof of this is given by Zhang {\em et al}.\cite{ZGR8836}.  The energy of this state is quite good.  The best estimate for the
ground state energy of the square lattice Heisenberg antiferromagnet which is
a N\'{e}el ordered state is $\langle S_{\bm i} \cdot S_{\bm j} \rangle = -0.3346$~J.\cite{TC8937,R9292}  The projected $\pi$ flux state
\cite{G8831} gives $-0.319$J, which is excellent
considering that there is no variational parameter.

In the presence of a hole, the dispersion of the staggered-flux phase is still given by Eq.~(12) with $\mu_f = 0$, but the Fermi level is now located at $\mu_f < 0$ which lies below the node.  This is shown in Fig.~7.  If the bosons are condensed, this becomes a Fermi liquid state with small hole pockets just like what is shown in Fig.~4(a).  This state has higher energy than the $d$-wave superconductor because in a superconductor the node is shifted in $\bm k$ space away from $(\pi/2a, \pi/2a)$ towards the origin, but its energy is always tied to the Fermi level.  The staggered flux state was proposed by Hsu, Marston and Affleck\cite{HMA9166} in 1991 to be the origin of the pseudogap state.  They pointed out that with doping, the staggered flux state has the remarkable property that orbital currents flow around each square plaquette in a staggered way, so that the physical order parameter of this state is the staggered orbital current.  This proposal did not receive much attention because the appearance of an ordered state with Ising order parameter requires a finite temperature phase transition, which has never been seen experimentally.  Furthermore, the model requires hole pockets instead of the arcs which were later observed.   While the matrix element effect can reduce the spectral weight of the backside,\cite{CNT0304} the model requires sharp quasiparticle peaks which should be observable, especially near the end of the arc, where the weight reduction is only a factor of 2.  In 2002, Chakravarty {\em et al.}\cite{CLM0203} revived this proposal, arguing that disorder effect may round the transition.  They named this state the  $d$-density wave (DDW) state.  Since this state is in fact identical to the staggered flux state of Hsu {\em et al.}\cite{HMA9166}, the two names are sometimes used interchangeably in the literature.  However, the philosophy of their approach is very different.  Chakravarty {\em et al.} take a phenomenological Landau theory approach.  For them the superconducting state is an entirely different order parameter and the two orders compete with each other.  I believe this view misses the key fact that all this is happening close to the Mott insulator.  While the $d$-wave superconductor and the staggered flux (DDW) states are very different states at the mean field level, the mean field states do not respect the no double occupation constraint.   The story changes if we enforce the constraint by applying the Gutzwiller projection.  As mentioned earlier, the projected states become identical at half filling and by continuity must share a lot of similarity slightly away from half filling.  For example, we found that at $x = 0.1$ the projected $d$-wave superconductor has short-range orbital current order\cite{ILW0058} and presumably the projected staggered flux state has short range $d$-wave superconductor order.  Both have significant short range AF order\cite{PRT0521}.  This kind of consideration motivated Wen and I to introduce the $SU(2)$ gauge theory in 1996 as an improvement over the Hsu {\em et al.} proposal.\cite{WL9603}  Instead of an ordered state, the pseudogap phase is considered to be a fluctuating state which included the staggered flux state and $d$-wave superconductivity on equal footing.  This will be described in more detail in the next section.

\begin{figure}[t]
\centerline{
\includegraphics[width=0.4\textwidth]{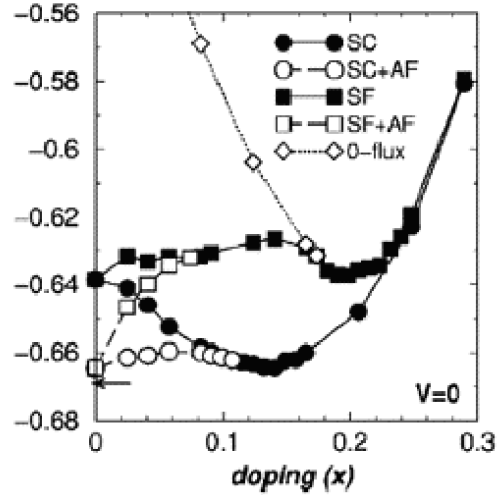}
}
\caption{ 
Energy per site of projected wavefunctions in units of $J$.  SF stands for staggered flux.  From Ivanov.\cite{I0403}}
\end{figure}

It is interesting to examine the energetics of various projected states as shown in Fig.~8.  It was found that the best state is a projected
$d$-wave superconductor and the sublattice magnetization is nonzero for $x  <
x_c$, where $x_c = 0.11$ for $t/J = 3$.\cite{I0403} The projected staggered-flux state always lies above
the projected $d$-wave superconductor, but the energy difference is small and
vanishes as $x$ goes to zero, as expected.  The staggered-flux state also
prefers antiferromagnetic order for small $x$, and the critical $x_c^{SF}$ is
now 0.08, less than that for the projected $d$-wave
superconductor.  The projected staggered flux state is the lowest energy
nonsuperconducting state that has been constructed so far.

\end{enumerate}

\section{SU(2), vortex core and theory of the pseudogap phase}

The lesson from examining projected wavefunctions is that the close relationship between different mean field states must be taken seriously.  In fact, the $d$-wave state and the staggered flux state are not the only states which have similar energies after projection.  There is a continuous family of states related by the $SU(2)$ symmetry given by Eq.~(14).  Wen and I developed a formulation to take this into account by introducing an $SU(2)$ doublet of bosons $(b_1, b_2)$ instead of the single boson in the $U(1)$ gauge theory.\cite{WL9603}  For our purpose this is just a technical way of generating different mean field states which are parametrized by rotating the quantization axis $\bm I$ in the $SU(2)$ space.\cite{LNN9803}  This way different states can be visualized and smoothly connected to each other in space and time.  In Fig.~9 we show such a representation.  The north and south poles correspond to the two degenerate staggered flux (DDW) states which break translational symmetry.  The equator is the $d$-wave superconducting state.  In between are many states which share both kinds of order.  One advantage of this approach is that we can construct a model of the vortex core of an $hc/2e$ vortex.\cite{LW0117}  This takes the form of a meron, or half a skyrmion.  As shown in Fig.~10 the center of the vortex is occupied by the staggered flux state.  This solves a serious problem with the original $U(1)$ formulation which favors the $hc/e$ vortex\cite{S9289,NL9266} because it is energetically favorable to retain the fermion pairing and make the boson wind by $2\pi$ around the vortex.  That approach ignored the fact that another state, the staggered flux state, is available to take the place of the normal state inside the core, allowing us to construct an $hc/2e$ vortex that costs very little energy.

Our model of the vortex core also explains a puzzling experimental observation, i.e. STM tunneling found that an energy gap remains when tunneling into the core region.\cite{PHG0036}  This is opposite to what was found in a conventional superconductor, where bound states are formed inside the core which fills in the gap.\cite{WM9576}

\begin{figure}[t]
\centerline{
\includegraphics[width=0.4\textwidth]{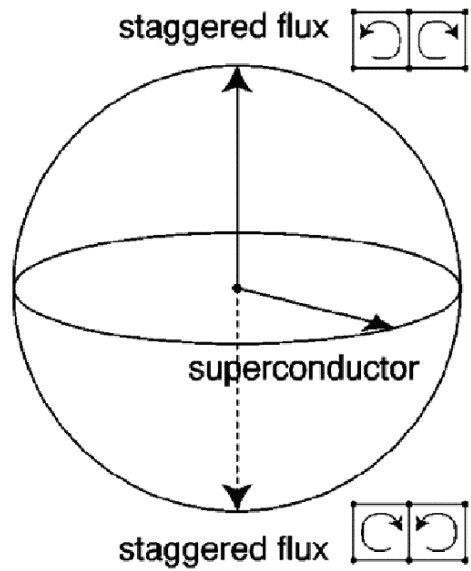}
}
\caption{The quantization axis $\bm{I}$ in the $SU(2)$ gauge theory.  The north
and south poles correspond to the staggered flux phases with shifted orbital
current patterns.  All points on the equators are equivalent and correspond to
the $d$-wave superconductor.  
}
\end{figure}

The $SU(2)$ construction allows us to construct a simple picture of the finite temperature phase diagram which accounts for the Nernst region and the pseudogap phase.\cite{LW0117}  This is shown in Fig.~(11).  At low temperature the $SU(2)$ quantization vector $\bm I$ is in the $x$-$y$ plane and we have the $d$-wave superconductor ground state.  As the temperature is raised, vortices are created with staggered flux cores and the proliferation of these vortices drives the Berezinskii-Kosterlitz-Thouless transition in the standard way.  Above this transition (which is the true superconducting $T_c$), vortex and anti-vortex proliferates giving rise to the Nernst phase.  At even higher temperature the $\bm I$ vector is completely disordered and this is our picture of the pseudogap phase.

We emphasized that in this theory the pseudogap phenomenon and superconductivity are intimately connected.  There is no separate pairing mechanism for superconductivity.  What drives superconductivity is the coherence of the boson to select the true ground state out of a myriad of fluctuating possibilities.  This is very different from the competing order scenario which does require a separate pairing mechanism and a completely separate energy gap scale.  
This dichotomy  has spurred a debate concerning one gap vs. two gaps, i.e. whether a smaller energy gap appears which scales with $T_c$.\cite{HHD0700}  
The debate is often couched in a black and white language, with one gap favoring a superconducting gap destroyed by phase fluctuations, and two gaps implying the need for some kind of competing order.  In my view, the truth is likely to be more complicated.
In the mean field RVB picture, a sharp quasiparticle peak appears below $T_c$ with weight $x$ which follows the $d$-wave dispersion with only one gap.  The mean field picture is probably too simplistic.  For example, it is possible that the large gap at $(0,\pi)$ as a spin gap which can remain broad while the low energy quasiparticles near the nodes become coherent below $T_c$.  To a first approximation, the coherent nodal quasiparticle has a dispersion which extrapolates to the large pseudogap at $(0,\pi)$.  There could be a coherence energy scale which scales with $T_c$, but exactly how this coherence scale affects the density of states and how it develops as a function of temperature is an open question.

The issue of one gap/two gaps is not
settled experimentally.  For moderately underdoped Bi2212 ($T_c > 50$~K) the evidence from ARPES is that the quasiparticle in the superconducting state is reasonably peaked even near the antinodes and obey the $d$-wave dispersion with a single gap which increases with decreasing doping.  This is supported by low temperature thermal conductivity data which measures the ratio $v_F/v_\Delta$ where $v_\Delta$ is the quasiparticle velocity in the direction of $(0,\pi)$.\cite{SHH0320} It is found that $v_\Delta$ increases with decreasing $x$ and extrapolates to the antinodal gap measured by ARPES.  On the other hand, for severely underdoped samples and one layer cuprates with low $T_c$'s, there are claims based on ARPES that the energy gap in the Fermi arc region near the nodal point does not scale with the pseudogap at $(0,\pi)$, which increases with decreasing doping as mentioned before.\cite{TAN0610}  Instead, it seems to stay constant or increase with decreasing doping.  
It is argued that this reveals a new gap scale associated with superconductivity.  I should caution that deeply underdoped samples are known to be strongly disordered, and the disorder increases with reduced doping.  Furthermore, the lineshape remains very broad in the antinodal direction even in the superconducting state.  
Thus it is risky to draw strong conclusion from lineshapes without an understanding of disorder effects and of the lineshape.

\begin{figure}[t]
\centerline{
\includegraphics[width=0.44\textwidth]{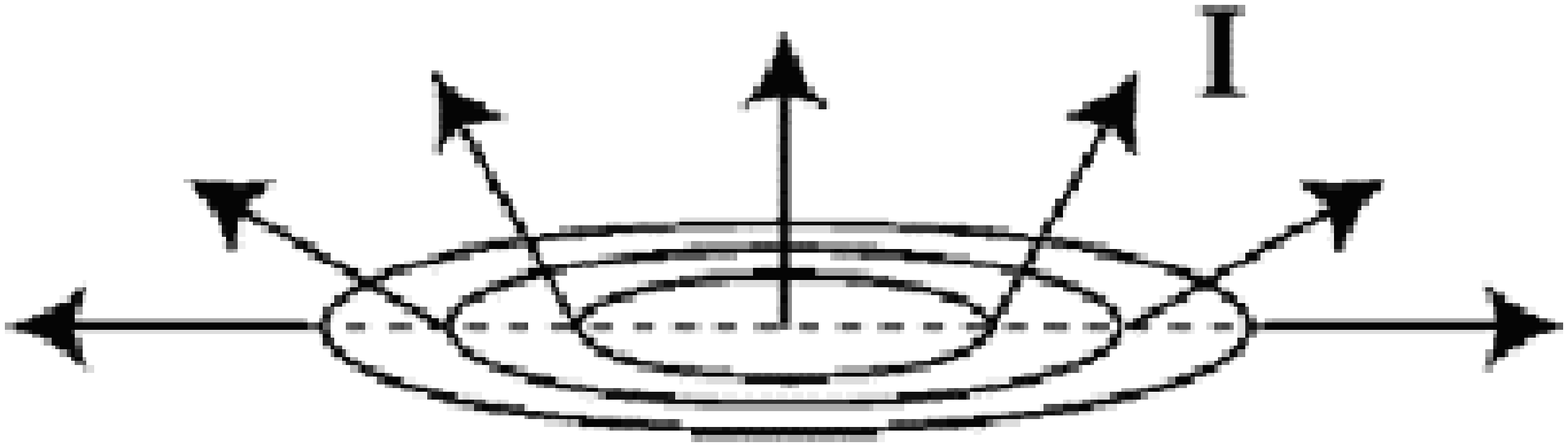}
}
\caption{ 
Model for a $hc/2e$ vortex.  The $SU(2)$ quantization axis points for the north pole at the center (forming a staggered flux vortex core) and rotates smoothly towards the equatorial plane as one moves out radially.}
\end{figure}

Other support for two gaps comes from Andreev reflection studies\cite{D9910} and Raman scattering.\cite{LET0637}  In a superconductor-normal metal junction in conventional superconductors, normal electrons incident on the junction has an extra channel for transport, by tunnelling as a Cooper pair into the superconductor and Andreev reflected as a hole.  This leads additional conductance below an energy scale of the energy gap.  Such extra conductance was observed in underdoped cuprates, but the energy scale observed is much lower than the pseudogap and is more related to $T_c$.  
I note that  in contrast to conventional tunnelling, Andreev reflection does not simply measure the density of states, but requires coherence of the quasiparticle in its interaction with the condensate.  What is seen in the Andreev data may be this coherence scale.

\begin{figure}[t]
\centerline{
\includegraphics[width=0.5\textwidth,height=0.33\textwidth]{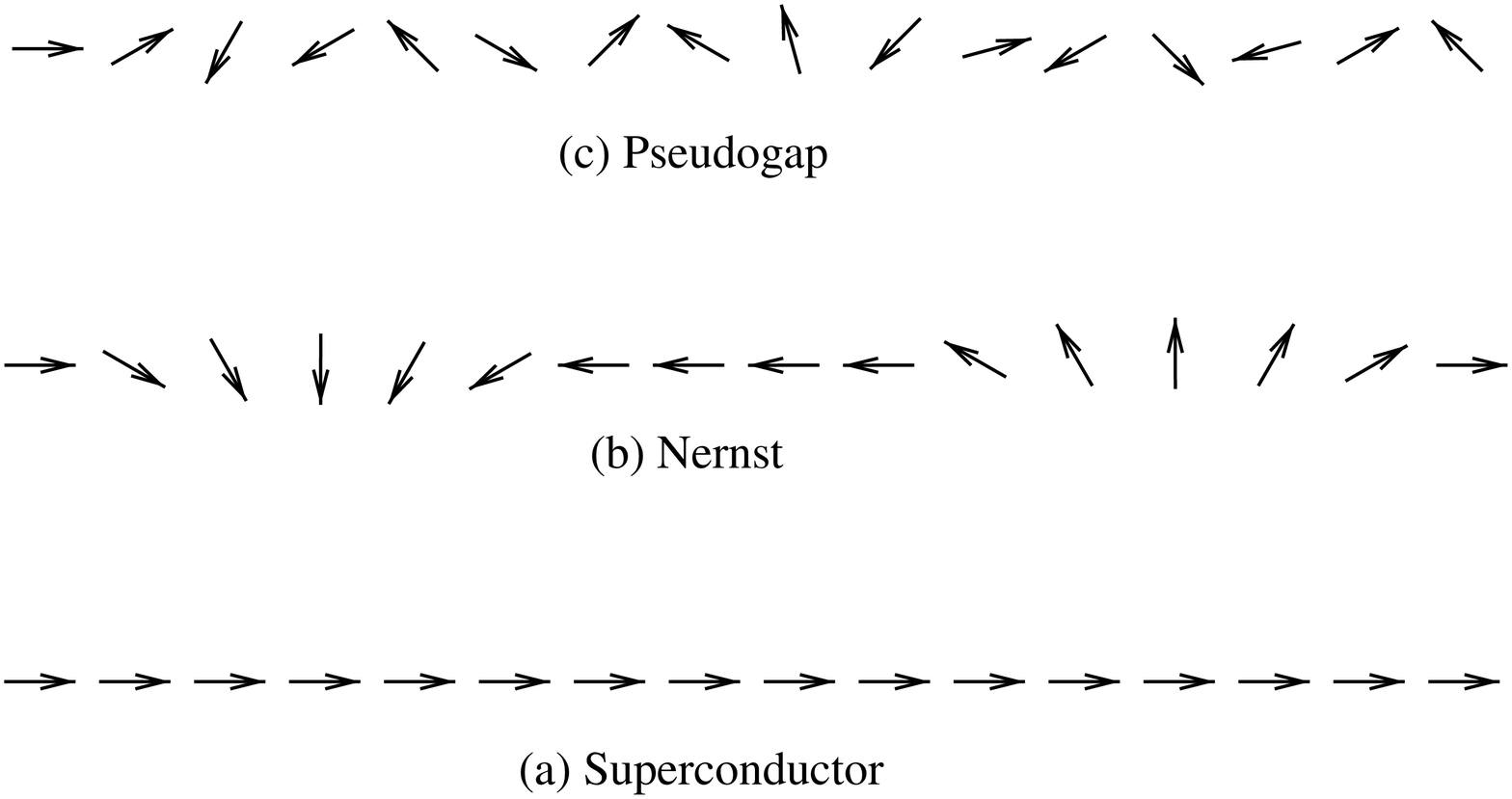}
}
\caption{ 
Schematic picture of the quantization axis $\bm{I}$ in different parts of the
phase diagram shown in Fig.~2. (a)~In the superconducting phase
$\bm{I}$ is ordered in the $x$-$y$ plane.  (b)~In the Nernst phase, $\bm{I}$
points to the north or south pole inside the vortex core. (c)~The pseudogap
corresponds to a completely disordered arrangement of $\bm{I}$. ($\bm{I}$ is a
three dimensional vector and only a two dimensional projection is shown.)}
\end{figure}

I must emphasize that the simple cartoon shown in Fig.~11 is only an approximate picture.  We have assumed that the bosons are locally condensed and can be treated as a $c$-number which varies in space and time.  
However, even at $T=0$, the vortex configurations shown in Fig.~10 can tunnel between each other and destroy the staggered flux order at some time scale.
I think the correct answer requires a quantum mechanical treatment of the boson strongly coupled to gauge fields, which is not available at present.  In particular, we have not yet been able to compute the ARPES spectrum and make a satisfactory comparison with experiment.  We make crude approximations such as assuming a binding of the bosons with fermions via gauge fluctuations.\cite{LNN9803}  As an example of an alternative aproach, Ribeiro and Wen\cite{RW0501} introduced a new formulation which hybridizes the physical hole with the spin carrying fermions and have had success in understanding the higher energy spectra.  Their theory seems to favor the two gap scenario.    The truth is that the theory of the spectral function is not under control at present: the problem of a fermion and boson strongly interacting with a gauge field at finite temperatures is too daunting with currently known techniques.  Rather than trying to explain data in detail, it may be more fruitful to step back and attempt a classification of the pseudogap state.  We have developed a view that the pseudogap phase belongs to the deconfined side of the lattice gauge theory, so that discussion in terms of fermions, bosons and {\em noncompact} gauge fields makes sense.  More explicitly, we propose that the pseudogap is best understood as the doping of a particular spin liquid called the algebraic spin liquid.  This is explained in the next section.

\section{Gauge theory, de-confinement and spin liquids}

Let us return to the gauge theory formulation described in section III.  As it stood Eq.~(8) is pretty much an exact reformulation of the $t$-$J$ model.  The constraint is enforced exactly upon integrating out the gauge fields.  However, the gauge field is fluctuating strongly because there is no restoring force, and critics of the gauge theory approach have pointed to this as evidence that this approach cannot be trusted to give meaningful results.  At a deeper level, I think what lies behind this objection is the implicit assumption that a phenomenon called confinement, familiar in QCD, necessarily takes place when the gauge coupling is large.  In this case, fermions are tightly bound to bosons by the gauge field, and one just recovers the original intractable electron problem with constraint.  In contrast, our way of thinking assumes that confinement does not happen, in which case the fermions, bosons and gauge fields retain their identity, albeit with strong interaction.  The ultimate strongly coupled fixed point is nontrivial and different from that of free particles, but as is usual in the strong coupling problem, it may be possible to access their behavior using artificial expansion parameters such as ${1\over N}$ where $N$ is the number of copies of the matter field.  Even though the physical problem may correspond to $N \sim$ 2 or 4, 
as long as confinement does not take place
the behavior of the physical system is smoothly connected to the large $N$ system and the physical behavior of the system can be understood, even though the critical exponents cannot be computed quantitatively.  

In the case of a compact $U(1)$ gauge, it is well known that in the 2+1 dimension the theory is confining even for arbitrarily small coupling, due to the proliferation of instantons. The point is that in a compact gauge field, a $2\pi$ flux can pop up through an elementary plaquette.  The space-time point where this happens is called an instanton event, which is also called a magnetic monopole in Euclidean space-time.  If these monopoles proliferate, the total flux through the sample is not conserved.  The strongly fluctuating gauge flux leads to confinement between external charges.  At first sight, this bodes poorly for the gauge theory approach.  However, the existence 
of matter field can change the story completely.  There is considerable experience in the study of a gauge field coupled to relativistic fermions and bosons, but little is known in the non-relativistic case.  In the past few years there has been significant progress in the case of half-filling where there are no bosons.  Consider the saddle point with $\pi$ flux through a plaquette.  The fermionic spectrum are massless Dirac fermions with nodes at $\left(  \pm {\pi \over 2a}, \pm {\pi \over 2a} \right)$.  Due to the additional $SU(2)$ symmetry, these fermions are minimaly coupled to a set of $SU(2)$ gauge fields. If the coupling to the gauge field is confining, this leads to what is known in QCD as mass generation and chiral symmetry breaking, which translates to gap formation and AF order in our language.\cite{KL9930} Thus the $\pi$ flux phase is our route to AF order.  
On the other hand, consider the staggered flux state.  This mean field ansatz breaks the $SU(2)$ symmetry and the gauge field is broken down from $SU(2)$ to $U(1)$.  This motivates the study of th problem of $N$ 2-component massless Dirac fermions coupled to a $U(1)$ gauge field in 2+1 dimensions.  This model is often called QED$_3$ in the literature.  Since the staggered flux state has two spins and two nodes, the physical problem corresponds to $N=4$.  (It was shown that the velocity anisotropy of the Dirac cone is an irrelevant variable and the low energy physics scales to the isotropic fixed or the QED$_3$ model.\cite{VTF0211,Lee/Herbut})
Assuming deconfinement, this class of states has been studied by ${1\over N}$ expansion and is called the algebraic spin liquid.\cite{RW0201}  Hermele {\em et al.}\cite{HSF0437} showed, using results  borrowed from the field theory literature, that for sufficiently large $N$, the deconfined state can be stable.  In other words, the large $N$ fixed point has no relevant perturbation, including the appearance of monopoles.  This is at least a proof of the principle that deconfinement is possible in the presence of matter field.  It is not known whether the critical $N$ is greater than or smaller than 4, and here is where the QCD Monte Carlo community can help.\cite{Fiore}  

Recently we received some encouraging news from experiments.  It has long been thought that the Kagome lattice with $S = {1\over 2}$ has sufficient frustration to support a spin liquid ground state.  This seems to be realized by recent experiments 
on ZuCu$_3$(OH)$_6$Cl$_2$  where the Cu ions occupy Kagome sites and the system shows no AF order down to 30~mK, despite an AF exchange of $\sim 200$~K.\cite{HEL0704}  On the theoretical side, it is found that the projected wavefunction of a certain flux state gives the ground state energy obtained by exact diagonalization to within error.\cite{RHL0705}  This is remarkable for a trial wavefunction with no variation parameter, and is much better than what projected flux states did for the square lattice.  The low energy physics of this state is that of two 2-component massless Dirac fermions with spins, coupled to a $U(1)$ gauge field.  The confirmation of this picture for the Kagome lattice will give us reason to believe that the critical $N$ is less than 4 in this case and perhaps in the case of the algebraic spin liquid as well.

Prior to the Kagome example, there was strong evidence that a spin liquid state exists in the organic compound $\kappa$-(ET)$_2$Cu$_2$(CN)$_3$.\cite{SMK0301}  Here the active sites are organic molecular dimer molecules which form a triangular lattice to a good approximation.  As mentioned in section III, the Heisenberg Hamiltonian on a triangular lattice is expected to order.  Nevertheless, experimentally no AF is found down to 32~mK.  The explanation lies in the fact that this system sits just on the insulating side of the Mott transition, so that charge fluctuations and ring exchange terms lead to a more complicated effective Hamiltonian which may favor the spin liquid state.\cite{M0505,LL0503}  
Numerical work has given an adequate account of the phase diagram, including the appearance of superconductivity under pressure as shown in Fig.~3.\cite{Kyung,Watanabe}
On the triangular lattice the low energy Lagrangian is expected to be an almost circular Fermi sea of spinons (fermions which carry $S={1\over 2}$ but no charge) coupled to $U(1)$ gauge fields.  This model has even more low lying fermionic excitations than the Dirac sea, and there is reason to believe that deconfinement occurs.  Experimentally the observations of a constant spin susceptibility and linear specific heat at low temperatures are highly unusual for an insulator, and 
support the notion of a spinon Fermi surface. 

The appearance of particles, such as spinons which carry different quantum numbers compared with the original electron, is a phenomenon called ``fractionalization.''  It is remarkable that the fermions originally introduced as a formal device in Eq.~(4) take on a life of their own at the end of the day.  Many people find this a hard concept to swallow and here is where explicit examples of a spin liquid will be a great help in convincing skeptics.  I must emphasize that the new structure in the theory (spinons, gauge fields, etc.) emerges in the low energy physics and what emerges is independent of the way the problem was formulated.  The constraint of no double occupation could have been enforced using $Z(2)$, $U(1)$ or $SU(2)$ gauge fields, but the low energy structure will be the same.  One particular formulation may simply be the convenient way to expose this emergent structure.

The discovery of experimental examples of spin liquids is a very important development because for the first time, the deconfined (fractionalized) states are low temperature states which in principle can be studied in great detail.  The success of the gauge theory method in these materials will give us confidence in its application to the more complex problem with holes.  It is fortunate that the two promising examples of spin liquids are closely related to those discussed in the high $T_c$ context for underdoped and optimal doping, respectively.

Finally, I would like to mention an approach to the pseudogap problem from a more general perspective.  Senthil and I \cite{SL0515} proposed that the underdoped cuprate should be considered as proximate to a spin liquid state which is then doped.  In this view the pseudogap is the finite temperature region controlled by the quantum critical point of doping a spin liquid by varying the chemical potential.  The most promising spin liquid is in fact the algebraic spin liquid with its low energy massless Dirac fermion and $U(1)$ gauge field. This point of view does not help solve the doped spin liquid problem, but does raise and answer the question: what is the signature of a deconfined state.  The answer is that in the presence of a matter field, the only signature left is the irrelevance of instantons, i.e. the total gauge flux is conserved, much like the conservation of magnetic flux in our world.  The experimental implication is subtle, but can be probed at least in principle in a suggested experiment.\cite{SL0515}

\section{Quantum oscillations in high magnetic field}

Is there a region of the phase diagram where the picture is less complicated and precise predictions can be made and tested?  One natural idea is to apply a large magnetic field perpendicular to the layer to kill the superconductivity and ask what is the next most stable ground state which emerges.  When $H$ reaches $H_{c2}$ the vortex core overlaps and the core state becomes the homogeneous ground state.  
Recent experiments indicate that we may be probing this regime and have caused a lot of excitement.\cite{DN0765,YEL0700}
It is believed that $H_{c2}$ for underdoped cuprates is approximately 100~T, well beyond what is available in the Laboratory.  The resistive transition, on the other hand, is controlled by flux flow and is much more accessible to experiment.  Recently, Shubnikov-de Haas oscillations are reported in an underdoped YBCO compound ($T_c = 55$~K) in the flux flow region, in a magnetic field range of 40~T to 62~T.  From the period of oscillation of the resistivity and Hall resistivity vs. ${1\over B}$ the area of the Fermi surface pocket is extracted and found to contain 0.038 holes per pocket.  If we assume  the doubled unit cell scenario, there are two pockets in the reduced BZ and this corresponds to $x = 0.076$.  The effective mass can also be measured from the temperature dependence, and $m^\ast = 1.9$ $m_e$.  The doping concentration of YBCO is difficult to estimate, because the doping is from the oxygen chains and its charge state is not independently measurable.  The best estimate by the experimentalists places $x$ at 10\%.  More recently, similar quantum oscillations are reported in YBa$_2$Cu$_4$O$_8$, the so-called Y124 compound ($T_c \sim 80$~K) with magnetic field from 55~T to 85~T \cite{YEL0700} and Shubnikov-de-Haas oscillations were reported in the range 45~T to 61~T.\cite{Bangura} The hole concentration per pocket is now 0.05 and the inferred $x$ of 0.1 is again lower than the value of 0.125 that is believed to characterize this material.  The effective mass is now $m^\ast = $ 3 $m_e$.

Quantum oscillation is considered the best way to measure the Fermi surface area, provided the sample is sufficiently free of defects that the electron can complete a cyclotron orbit without being scattered.   Even at 50~T, the cyclotron orbit circumference is more than a thousand $\rm{\AA}$ in size (see Fig.~12), thus making it clear that disorder is really not an issue in the YBCO family.

 \begin{figure}[t]
\centerline{
\includegraphics[width=0.45\textwidth]{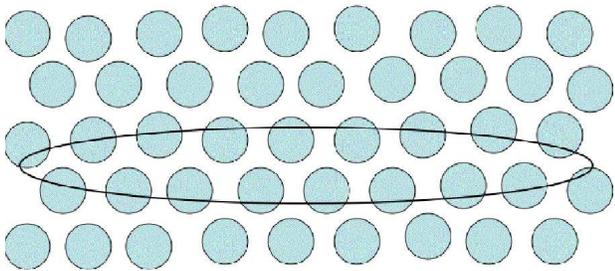}
}
\caption{
Vortex cores are represented by discs of radius $R_v$.  Figure is drawn almost to scale at $H={1\over 2}H_{c2}$, when the discs occupy half of the total area ($R_v\approx 25\rm{\AA}$).  A cyclotron orbit is shown, also roughly to scale.  Its area encloses about 10 vortices for the experiment on YBCO.  The long dimension of the orbit is about $800 \rm{\AA}$.}
\end{figure}

I think there is an excellent chance that the experiment is accessing the normal state that lies beyond $H_{c2}$. It is well known experimentally that quantum oscillations persist to $H < H_{c2}$.\cite{JAN9898}  The frequency of the oscillation is not changed upon crossing $H_{c2}$, only its magnitude is diminished.  This phenomenon is particularly striking in layered materials where the flux flow regime can be very wide.  For example, a recent experiment on organics observed quantum oscillations down to ${1\over 2} H_{c2}$ and is even more striking in that the expected reduction of the amplitude was not even observed.\cite{WOS0073}  A rough physical explanation may be as follows.  Let us take a cartoon picture of the vortex core as discs of radius $R_v$.  At $H_{c2}$ the discs overlap and we define $H_{c2} = \tilde{\phi}_0/(\pi R_v^2)$ where $\tilde{\phi}_0 = hc/2e$ is the superconducting flux quantum.  Note $R_v = \sqrt{2}\xi$ where $\xi$ is the coherence length.  For $H_{c2} = 100$~T, $R_v $ = 25~$\rm{\AA}$.  For $H = {1\over 2} H_{c2}$ the disc density is rather high, as drawn in Fig.~12, almost to scale.  Let us estimate the size of the semiclassical orbit.  Very generally, the real space area $\tilde{A}$ is related to the momentum space pocket size $A_k$ by

\begin{equation}
\tilde{A} = A_k \left(
{hc\over eH}
\right)^2  .
\end{equation}
It is useful to introduce the Landau length 

\begin{equation}
\ell_H = \sqrt{{hc\over eH}}  .
\end{equation}
Note that the flux through an area $\pi \ell^2_H$ is $\tilde{\phi}_0$, so that $R_v = \ell_{H=H_{c2}}$.  The Landau level number $\nu$ at the Fermi level is given by the number of full flux quantum $\phi_0 = 2\tilde{\phi}_0$ which penetrates $\tilde{A}$, i.e.

\begin{equation}
\nu = {\tilde{A} \over 2 \pi \ell_H^2} = {A_k \tilde{\phi}_0 \over 2\pi^2 H} .
\end{equation}
We find $\nu = 10$ for $H = 50$~T, using the experimentally measured $A_k = 5.1 \times10^{14}$~cm$^{-2}$ for YBCO.  Since 50~T corresponds to $\approx {1\over 2} H_{c2}$, this also means that on average, 10 vortex cores are inside the spatial orbit.  Since the cyclotron motion is fast compared with that of the vortices, we can assume a static picture of the vortex core, which either forms a hexagonal lattice or somewhat distorted from it.  Assuming the orbit to be an ellipse with aspect ratio $\sim 8$ (for reasons to be explained later), we sketch a snapshot of the orbit in relation to the vortex core in Fig.~12.  In order to see quantum oscillations, what is required is that the states in the cores are in phase, whether it is AF order or staggered-flux order.  In the latter case, the $\bm I$ vector in all the vortices inside the cyclotron orbit are either pointing up or pointing down, i.e. the coherence length of the order must exceed the orbit size.  For the aspect ratio chosen, the long dimension of the orbit is about 800~$\rm{\AA}$.  What is required for quantum oscillations is strong tunnelling between the vortex cores.  Then the quasiparticles are extended states which carry information of the uniform normal state for $H > H_{c2}$ because the random phase coming from tunnelling through the superconducting region will cancel.  Given the dense packing of vortices, this seems quite reasonable.

What is the origin of the pockets that give rise to the quantum oscillations?  Here we are on rather uncertain grounds.  In general, we can classify the state as conventional or exotic. By conventional I mean there is some new order giving rise to unit cell doubling and a Fermi liquid ground state, and exotic means everything else.  While exotic scenarios have been suggested \cite{KKS0700} here I will concentrate on discussing conventional possibilities, knowing that we immediately encounter a problem concerning the relationship of the measured area to doping estimates.  Either the doping estimate is incorrect, or there exists undetected electron or hole pockets.  Let me forge ahead and assume that the doping concentrations are actually $x=0.075$ and 0.10 and correspond to the pocket areas measured in the two experiments and discuss several candidates for the unit cell doubling order. 

\begin{enumerate}

\item AF order.  We saw in section I that at very low doping, hole pockets appear around $(\pi , \pi)$ in the background of AF order.  The question is whether this narrow region $(x < 3\%)$ will open up in a strong magnetic field and extend up to 10 or 12\% doping.  This is probably what most people have in mind when they refer to AF in the core and we shall call states of this kind ``doped AF.''  I think this scenario is unlikely.  From neutron scattering we know that there is a sharp triplet resonance at $\sim 30$~meV in this doping range at $(\pi , \pi)$ and a sharp drop in spectral weight below it.\cite{Stock}  AF order requires the condensation of this triplet mode.  A 50~T field corresponds to an energy scale of 6~meV, hardly enough to perturb the resonance mode.  Thus I think the AF is energetically highly unfavorable.  Nevertheless, there are experimental ways to distinguish doped AF from the staggered flux state, which I will return to later.

\item Staggered flux state.  As we argued before, this state is energetically favored to appear once superconductivity is destroyed.  I want to point out that in principle a small amount of AF order can co-exist with this state.  As seen in Fig.~8, this is favorable for $x < 0.08$.  However the driving force and excitation spectrum of this state is completely different from the doped AF.  In that case the holes live in the lower Hubbard band, i.e. the band is a downward parabola, separated by a gap of order of the Mott gap with energy scale 1~eV.

\item Valence bond solid (VBS), or more general states which break translation symmetry without breaking time reversal as in AF.  There has been discussion of this state living in the vortex core.\cite{SO313}  Roughly speaking, the electronic spectrum may be quite similar to the staggered flux state except that it is expected to be gapped.

\item Spin density wave (SDW).\cite{CYR0700}  This seems very attractive phenomenologically.  From NMR there are claims for enhanced antiferromagnetism in the core.\cite{MIT0303,KKM0303} In the doped La$_2$CuO$_4$ system it is known from neutron scattering that incommensurate SDW  is induced in the vicinity of the vortex core.\cite{Lake,KLE0228}  This is quite understandable because inelastic neutron scattering indicates soft excitations at precisely these incommensurate wave vectors, typically around ${2\pi\over 8a}$ from $\left(  {\pi\over a}, {\pi\over a} \right)$.  This is often connected to the notion of stripes near the concentration $x = {1\over 8}$.  However, static SDW has never been seen in the YBCO family.  Neutron scattering has reported dynamical fluctuations at 20~meV at the incommensurate wave vectors $\left( {\pi\over 2a} \pm \delta, {\pi\over 2a} \pm\delta \right)$.  While recent data on more underdoped samples ($T_c = 50$~K) sees dyamic scattering down to 12~meV, it has much less spectral weight than in the doped La$_2$CuO$_4$ system.\cite{Stock}  Thus it is questionable whether a magnetic field can induce SDW order.  Furthermore, the prevailing view is that the SDW is not tied to Fermi surface nesting, and it is unclear whether hole pockets should be expected.

 \begin{figure}[t]
\centerline{
\includegraphics[width=0.4\textwidth]{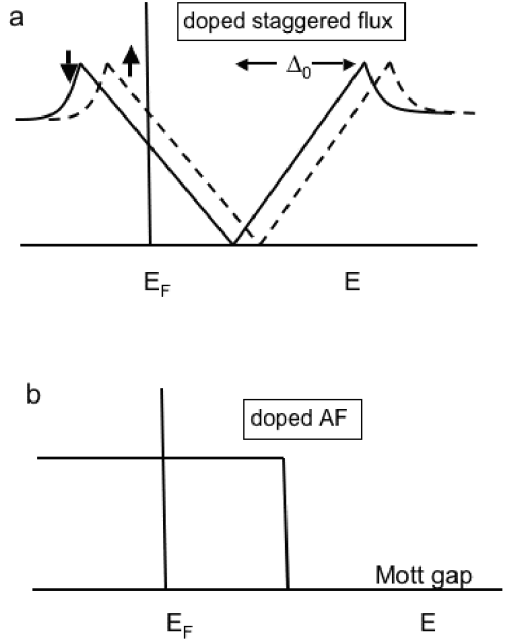}
}
\caption{ 
Schematic picture of the tunneling density of states for (a) the spin split bands in the doped staggered flux state and (b) the doped AF.  If the hole pocket lies in the lower Hubbard model, a large Mot gap is expected.}
\end{figure}

How can we distinguish the staggered flux state from the doped AF state? A useful experimental test is to do a tunnelling experiment to measure the density of states in high fields.  Since spatial resolution is not needed, we do not need STM, which is probably impossible under these pulsed field conditions, but we propose either break-junction tunnelling or other tunnelling geometry where electrons can be injected into the a-b plane from the side.  Schematically the density of states of the staggered flux state vs. the doped AF is sketched in Fig.~13(a).  Of course, we expect much of the higher energy features to be smeared out, but strong particle-hole asymmetry and a dip in the density of states above $\varepsilon_F$ will support the staggered flux picture.  On the other hand, if this state is the continuation of the doped AF state observed at $x < 0.03$ for zero field, we expect a large Mott gap above the Fermi energy, which has been estimated to be 30~meV in the case of Y124.\cite{YEL0700}  Thus tunnelling data is a sensitive test of the notion of hole pockets in the lower Hubbard band.

Next we extract some physical parameters from the data.  Let us parametrize the Dirac spectrum of the staggered flux state near each node by

\begin{equation}
E(k) = \sqrt{(v_Fk_1)^2 + (v_\Delta k_2)^2}
\end{equation}
where $k_1(k_2)$ in the direction perpendicular (parallel) to the large Fermi surface.  By definition

\begin{equation}
m^\ast = {\hbar^2 \over 2\pi} {dA_k \over dE} .
\end{equation}
By writing the density of states $m^\ast/2\pi$ in terms of the velocities along the Fermi surface, we find the anisotropic generalization of the relation $v_F = \hbar k_F/m$ :

\begin{equation}
\sqrt{v_Fv_\Delta} = {\hbar \over m^\ast}\sqrt{{A_k \over \pi}} .
\end{equation}
Equation~(20) applies equally well to the anisotropic Dirac spectrum and the anisotropic parabolic band.  The right hand side of this equation are directly measured.  Putting in the numbers for YBCO with $m^\ast$~=~1.9~$m_e$, we find $\hbar\sqrt{v_Fv_\Delta}$ to be 0.48~eV$\rm{\AA}$.  The Fermi velocity can be measured by ARPES, even though the error bar is substantial.  Taking the value $v_F$~=~1.4~eV$\rm{\AA}$ from Campuzano {\em et al.},\cite{CNR03} we find the anisotropy to be $v_F/v_\Delta = 8.4$.  This compares with the value 7.9 directly extracted from thermal conductivity data \cite{SHH0320} in their $T_c = 62$ sample.  In that paper they extract a gap value $\Delta_0$ of 71~meV, using $v_F$~=~1.65~eV$\rm{\AA}$.  With our numbers we expect $\Delta_0 = 57$~meV, in very good agreement with ARPES estimate of the pseudogap for a similarly doped Bi2212 sample.  A similar exercise using Y124 data \cite{YEL0700} yields $v_F/v_\Delta = 16.7$ and $\Delta_0 = 29$~meV.  The gap value is on the low side.  Given that these estimates are sensitive to the square of $m^\ast$ and $v_F$ and their associated errors, these are reasonable numbers.  More importantly, it shows the correct trend of increasing anisotropy and reduced gap with increasing doping.  It is worth noting that the ellipse drawn in Fig.~1(b) by Doiron-Leyraud {\em et al.}\cite{DN0765} has an aspect ratio of $v_F/v_\Delta \approx 4$.  Noting that the length of the ellipse scale as $\sqrt{{v_F\over v_\Delta}}$, our numbers indicate a more elongated ellipse and a small departure of the inside edge of the pocket away from the Fermi arc.  For Y124, the ellipse almost reaches the saddle point at $(0,\pi)$.  Beyond this point the pockets connect to form large Fermi surfaces which are presumably unobservable due to disorder scattering, and we expect the quantum oscillation to disappear.  There will be a transition to the uniform large Fermi surface (area $1-x$) state, but the exact nature of this transition is an open issue.  We reiterate that Eq.~(20) applies to the parabolic band of the doped AF as well.  The only difference is that in that case there is no reason to identify $v_F$ with the normal state Fermi velocity.

Quite generally, if this state is a Fermi liquid in a nonmagnetic background, we expect Zeeman splitting of the up- and down-spin pockets.  For free fermions the splitting is entirely determined by $m^\ast$ which gives the density of states.  Let us define $\Delta A_k = A_{k\uparrow} - A_{k\downarrow}$ to the difference in the up- and down-spin pockets.  We find

\begin{equation}
\Delta A_k = \Delta E {dA\over dE} = \Delta E 2\pi m^\ast/\hbar^2
\end{equation}
Using $\Delta E = g\mu_BH$ with $g=2$ and $H=60$~T, we find for Y124 a splitting of $\Delta A_k/A_k \approx 26\%$.  Experimentally a splitting of $\approx 10\%$ was observed.  If we attribute this to spin splitting, we will need a reduction of the predicted magnetization due to residual interactions.  In Fermi liquid theory, this is given by the Landau parameter $F_0^a$, so that $\chi = \chi_0/\left( 1 + F_0^a \right)$.  Thus we need $F_0^a = 1.6$ which is quite reasonable for a strongly interacting system.

In contrast, for the doped AF, we do not expect a similar spin splitting. This was pointed out to me by T. Senthil.  This is because the AF will cant with sublattice magnetization perpendicular to the applied field and spin along the field is no longer a good quantum number.  If splitting due to interlayer tunnelling can be ruled out, this is another argument against the doped AF.

Finally, we discuss the cyclotron resonance experiment.  In principle, this can help distinguish between the Dirac spectrum of the staggered flux state and the parabolic spectrum expected for the doped AF.\cite{LW0117}  As made famous by recent experiments on graphene,\cite{NOV0597,Zha0501} the Landau level in a Dirac spectrum is given by $E_\nu = \sqrt{2\nu} \hbar\sqrt{v_Fv_\Delta}/\ell_{BH}$ which scales as $\sqrt{B}$, instead of the standard Landau level spacing $\hbar eB/m^\ast c$ which is linear in $B$ in conventional parabolic bands.  Using measured parameters for YBCO we find the splitting between the $\nu = 9$ and $\nu = 10$ Landau level to be 3.1~meV, compared with 3.3~meV for a parabolic band, an unfortunate coincidence.  In principle, one can check the $\sqrt{B}$ vs. $B$ dependence, but this may be a challenging experiment to carry out in pulsed fields.

\begin{figure}[t]
\centerline{
\includegraphics[width=0.5\textwidth]{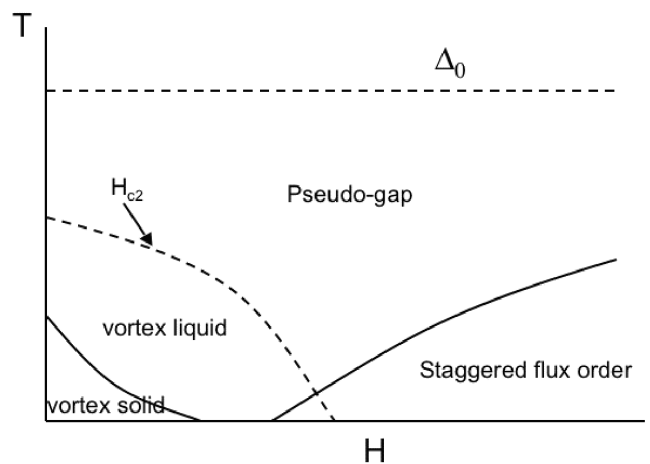}
}
\caption{
Schematic phase diagram predicted for an underdoped cuprate in a magnetic field.  The solid lines are true phase transitions while the dashed lines are cross-overs to the pseudogap phase and the vortex liquid phase.  The vortex liquid at $H=0$ corresponds to the Nernst region in Fig.~2.}
\end{figure}

To summarize, with insight gained from Nernst effect measurements, we propose in Fig.~14 a phase diagram for the underdoped cuprate in a field.  The staggered flux state appears as a new phase in high field and low temperature.  Since this state requires the phase coherence of the bosons just as the superconducting state, we expect its transition temperature to be comparable to the superconducting $T_c$.  We expect a signature in transport properties at this transition temperature.  The $H_{c2}$ line is  a cross-over line which characterizes the onset of a vortex liquid state.  This is the extension of the Nernst region shown in Fig.~2 to finite $H$.  The true superconducting transition is to a vortex solid or glass state.  Whether the staggered flux state terminates in the vortex liquid or solid phase is not known.  Below a large energy scale $\Delta_0$ lies the spin gap phase which is the extension of the pseudogap phase in Fig.~2 to finite $H$.    We emphasize that while the model of Hsu {\em et al.}\cite{HMA9166} and the more phenomenological approach of Chakravarty {\em et al.}\cite{CLM0203} also predict a staggered flux (DDW) state at high field, they key point they made is that this state is smoothly connected to the pseudogap phase, i.e. the solid phase boundary in Fig.~14 is pushed up to an energy scale $\Delta_0$.  Great effort has been made to look for the moment generated by the orbital currents by neutron scattering and nothing convincing has been found.  There is no sign of a phase transition even in the highest purity materials.  It is difficult to see how the Fermi arc can be consistent with a Fermi liquid picture of sharp quasiparticles.  In my opinion, existing data at low field is sufficient to rule out the DDW scenario.  In contrast, in our picture the pseudogap phase is separated from the high field staggered flux Fermi liquid state by a phase boundary.  While a detailed description of the Fermi arc is not available, it is at least possible to imagine that different physics is at play and the Fermi arc picture can be consistent with the experimental observation of quantum oscillations in high fields.

Finally, the proposed phase diagram is heavily based on mean field theory and the notion of boson condensation and should be viewed simply as a starting point for discussion of the experiment.  It is possible that the orbital current can survive fluctuations, but it is also possible that tunnelling events wipe out the ordering pattern, leaving us with an exotic ground state.  Other order, such as weak AF, could also emerge to take its place.  The overall picture, I think, is more general than the survival of the orbital currents.  The large spin gap at $(0,\pi)$ and features in the tunnelling density of states sketched in Fig.~13(a) may survive.  Clearly more experiments in high fields will be needed to shed light on these important issues.
\end{enumerate}

\section{Questions and answers}

Instead of a conclusion, I would like to pose a number of questions and attempt to answer them.

\begin{enumerate}

\item What is unique about the cuprates?  What is the pairing mechanism?  What controls $T_c$?  Can $T_c$ be raised?

The cuprate is the prime example of a two-dimensional Mott insulator which can be doped.  The copper $2+$ ion is in a $d^9$ configuration, making it a spin ${1\over 2}$ system.  The layer symmetry breaks the degeneracy of the $d$ orbitals dow to a single $d_{x^2-y^2}$ orbital, so that orbital issues which cause complications in most other transition metal oxides do not play a role here.  In most other examples, strong coupling of the charge state to distortion of the oxygen cage causes doped carriers to localize.  For example, if we replace Cu by Ni, La$_2$NiO$_4$ requires 50\% hole doping before it becomes metallic.  In many ways, the cuprate is the ideal model system for a doped Hubbard model.  The parent compoud is one of the few layered $S = {1\over 2}$ antiferromagnet with no orbital degeneracy that is known.  It is hard to find a material that is so simple!

Searching for a pairing mechanism may be the wrong question, because implicitly the assumption is made that pairing occurs via the conventional route of well defined quasiparticles exchanging some bosonic glue.\cite{P0705}  We know that underdoped cuprates do not fit this description.  Instead superconductivity is the ground state which emerges as the best compromise between the kinetic energy of hole motion and the antiferromagnetic exchange.  The energy gap is set by the exchange energy $J$ while $T_c$ is controlled by phase coherence and is of order $xt$.  Thus the optimal $T_c$ is set by $J$.  The cuprate family already has the highest $J$ among transition metals.  (The ladder compound made up of the same copper-oxygen bond is the only example I know which has a higher $J$ than the planar cuprate.)  The hopping matrix element $t$ is also large because of the strong covalent bonding between the $d_{x^2-y^2}$ orbital with the oxygen $p$ orbital.  Thus the cuprates are blessed with a large energy scale.  The variation of $T_c$ on a few tens of degrees between YBCO and the Hg compound seems to be attributable to changes in the further neighbor hopping matrix elements.\cite{PDS0103}

If the superconductivity is due to the electronic energy scale such as $J$ and $t$, why isn't $T_c$ higher?  As we saw in section I, it is only $\sim {t\over 40}$ or ${J\over 10}$ for both cuprates and organics.  What accounts for the small numerical factor?  One consequence of doping a Mott insulator is that the carrier density and therefore the superfluid density is small.  Thus phase fluctuation is important and limits $T_c$.  In a $d$-wave superconductor, the thermal excitation of nodal quasiparticles further suppresses the superfluid stiffness on a short distance scale, as seen by experiment.\cite{CMO9921}  This reduces the $T_c$ further.  This is why $T_c$ is such a small fraction of $J$.  Besides increasing the overall energy scale, one possible avenue for enhancing $T_c$ is to try to stabilize superconductors with a full gap such as $d+id$ states.  For example, on a triangular lattice, the $d+id$ state is believed to be favored.  In this case, the quasiparticle suppression of the superfluid density is negligible and $T_c$ will be higher, everything else being equal.

\item What about other approaches?  Why is there no consensus after all these years?

In a review of this kind which adopts a particular viewpoint (going under the names strong correlation, RVB, gauge theory, etc.) I feel obligated to give my honest opinion on other approaches.  This may not be the best way to win friends, but let me proceed.  
First let me point out that within the RVB approach there are many variants that have not been covered here.  I emphasized the formulation based on a fermionic representation of the spins and a bosonic representation of the holes.  The opposite formulation with bosonic spinons  \cite{SO313} and a more complicated formulation with flux attachment \cite{W0773} has also been pursued.  Below let me focus on approaches outside the RVB ``big tent'' by separating the most prominent work into four major categories and discussing them in turn.

\begin{enumerate}

\item Spin fluctuation models.

Historically the possibility of $d$-wave superconductivity based on the exchange of AF spin fluctuations was discussed prior to the discovery of high $T_c$ and a number of workers apply these ideas to the cuprates.\cite{E8621,MSV8654,SLH8690,MP9369}  These discussions are either based on random phase approximation treatment for the Hubbard model, or phenomenological coupling between spins and electrons treated as separate entities.  The latter may be correct in heavy fermion systems, where there indeed exists separate itinerant and local moments, but is more questionable in the Hubbard model where there is only one kind of electronic state.  The best chance for this model to work is in the overdoped region, where well defined quasiparticles exist in the normal state and the onset of superconductivity appears quite conventional.  However, from neutron scattering it is known that spin fluctuations become very weak, with spin correlation length of 10~$\rm{\AA}$ or less beyond optimal doping.  A variant of this idea ascribes the boson being exchanged to the sharp triplet resonance discovered by neutron scattering.  However, this resonance is very narrow in momentum space and carries very little spectral weight, making it unlikely to be the powerful glue needed.\cite{KKA0202}  
It is clear that in order for the spin fluctuation picture to work, one has to deal with the strong coupling problem and include spin fluctuations at all energy scales up to $J$.  Numerical methods may be the only option.  Recent numerical studies based on cluster dynamical mean field theory shows promise in this direction in that a pairing kernel was identified in the triplet channel which peaks at $\left({\pi\over 2}, {\pi\over 2}\right)$ and resembles the spin fluctuation channel.\cite{Maier}
I think the most serious limitation of this line of ideas is that it completely fails to address the pseudogap issue.  As we have seen, there is simply no well defined quasiparticle above $T_c$ to pair them by whatever mechanism.  By avoiding the issue of proximity to a Mott insulator, this approach misses the key physics and the most interesting aspect of high $T_c$.

\item  Microscopic inhomogeneity, stripes, etc.

The basic idea is that one way the doped Mott insulator can resolve the competition between kinetic energy and AF exchange is for the holes to phase segregate.  In particular, the holes may concentrate into one dimensional regions where AF order is suppressed, separated by an AF region where the hole density is suppressed.\cite{CEK03}
Experimentally, it was discovered that a particularly stable configuration where the holes form a quarter filled chain (one hole per two sites along the chain) separating AF regions as an anti-phase boundary indeed exists in La$_{2-x}$Ba$_x$CuO$_4$ where $x = {1\over 8}$.\cite{TSA9561}  This configuration and variants thereof has been called stripes.  The stripe explains the suppression of $T_c$ in the doped La$_2$CuO$_4$ system near $x = {1\over 8}$.  It also receives theoretical support from numerical solution of the $t$-$J$ model using density matrix renormalization group method.\cite{WS9953}  Thus there is good reason to believe that stripes form a strongly competitive ground state, especially near $x = {1\over 8}$.  However, the ordering temperature for the charge order which preceded spin order 
is low, of order 40~K, and the spin order (incommensurate SDW) is even lower.  Furthermore, static stripes are seen only in the doped La$_2$CuO$_4$ family and not in the YBCO family, and in the latter case, even dynamical stripe fluctuations have weak intensities, as discussed earlier.  Thus it is my belief that stripes represent low energy physics and emerge as the ground state only as a result of delicate competition.  Recently there is a remarkable report that the ARPES in La$_{1.875}$Ba$_{.125}$CuO$_4$ reveals quasiparticles which $d$-wave dispersion with a maximum gap of 20~meV, as if the existence of stripes has no effect on pairing of the pseudogaps except at very low energy.\cite{VFL0614}  This reinforces my view that stripe order is a secondary effect.

Static stripes are clearly detrimental to superconductivity.  Nevertheless, there has been a strong push, notably by S. Kivelson and co-workers, for the idea that fluctuating stripes may be responsible for superconductivity and the pseudogap phenomenon.  I have always found the motivation of this idea puzzling.  Fortunately a recent article by Kivelson and Fradkin\cite{KF07} lays out the physical ideas and motivation in a very clear way.  They share our basic assumption that doping a Mott insulator is the key question and they share our notion that superconductivity comes from strong repulsion.  However, they reject the possibility that this can happen in a homogeneous phase.  As far as I can tell, the reason offered is that ``the enormous effort has been devoted to numerical searches for superconductivity in various uniform Hubbard and $t$-$J$ related models, with results that are, at least, ambiguous'' and the feeling that if that were the right route, ``unambiguous evidence of it would have been found by now.''  Instead they focused on models where superconductivity from a repulsive Hubbard model can be demonstrated, namely the ladder system.  The ladder system can be considered an example of a valence band solid, where spins on the rung form singlets and doped holes tend to form pairs on the rung.  Then pair hopping between ladders provides bulk superconductivity.  This is all well and good but Kivelson and Fradkin are the first to admit (in fact insist) that this toy model has little to do with cuprates.  For example, it is hard to imagine how quasiparticles can traverse this system or ladder (or any kind of stripes) at a $45^\circ$ angle and become the gapless nodal particle.  As long as this problem is not addressed adequately, I fail to see how this approach can be a viable route to superconductivity in the cuprates.

\item Phonons.

Electron phonon coupling is typically strong in transition metal oxides involving the $E_g$ orbitals, because of strong coupling to the distortion of the oxygen cage.  As mentioned earlier, in most cases this leads to localization of the doped hole.  While holes in the cuprate escape this fate, the electron phonon coupling must be strong.  Nevertheless, we have argued that the dominant energy in this problem is Coulomb repulsion.  Hence the discussion of electron-phonon coupling must be made in this context.  A combination of numerical and analytic work has shown that short range repulsion has the effect of reducing electron-phonon scattering at large momentum transfer because charge fluctuations at short distance is suppressed.\cite{MN0402,BRC0620}  It is also claimed that under certain circumstances, $d$-wave pairing is possible with electron-phonon coupling, contrary to conventional wisdom.  Experimentally, phonon sidebands appear visible in STM tunnelling spectra.  All this says that electron-phonon coupling is there (no great surprise) but does not provide any evidence that it is the essential driving force behind pairing.

The isotope effect is often quoted as evidence for the phonon mechanism for high $T_c$.  It turns out that isotope effects affect $T_c$ via the superfluid density, and not the order parameter itself.\cite{KSC0317}  Thus the isotope effect in fact supports the view that $T_c$ is controlled by phase fluctuations in the underdoped region.  There is no isotope effect at optimal doping and beyond.

\item Three band model.

Since the copper oxygen layer involves one copper and two oxygen per unit cell (excluding the apical oxygen), the minimal microscopic electronic model requires a $d_{x^2-y^2}$ Cu orbital and two oxygen $p$ orbitals.  This is called the three band model.\cite{E8759,VSA8781}  Various cluster calculations indicate that the low energy physics (below a scale of $t \approx 4$~meV) can be adequately described by a one band Hubbard model.\cite{ZR8859}  Over the years this has become the majority view, but there are still workers who believe that the three band model is required for superconductivity.  I think the original motivation for this view came from a period of nearly ten years when much of the community believed that the pairing symmetry is $s$-wave.  It certainly is true that a repulsive Hubbard model cannot have an $s$-wave superconducting ground state.  Three band models with further neighbor repulsion were introduced to generate the requisite effective attraction.  The $s$-wave story was overturned once and for all by phase sensitive experiments in 1994 but some of the three band model proponents persevered.  In particular, Varma introduced the idea of intra-cell orbital currents, i.e. a current pattern flowing between the Cu-O and O-O bonds as a model for the pseudogap.\cite{V9754}  This has the virtue of leaving the unit cell intact and this kind of order is very difficult to detect.  With such a complicated model it is difficult to make a convincing case based on theory alone and a lot of attention has been focused on experimental detection of time reversal symmetry breaking or spontaneous moments due to orbital currents.  Unfortunately, the expected signals are very small and can easily be contaminated by a small amount of minority phase with AF order.  Up to now there is no reliable results in support of this kind of orbital currents.
\end{enumerate}

\item 
Why is the high $T_c$ problem hard?  Why is the high $T_c$ problem important?

One of the ironies about the high $T_c$ problem is that the ground states are rather conventional.  It ranges from AF to $d$-wave superconductors to Fermi liquid as the doping is increased.  The exciting new physics happens at finite temperatures where Fermi arcs, pseudogaps and other novel phenomena appear.  This region is not characterized by order in the traditional Landau sense, and it is not possible to make precise statements as one could if one has an exotic ground state.  From the theoretical side, what we can do so far to classify the pseudogap as the deconfined phase of a gauge theory, where new particles, spinons and holons and $U(1)$ gauge fields emerge in the low energy physics.  These particles and gauge fields are still strongly coupled and we are not able to compute and compare with experiments in a quantitative way.  We can make various caricatures (such as local boson condensation and fluctuations between different phases) which mimic the observations in a qualitative manner.  Nevertheless, I believe the problem is so strongly constrained by data that this path is the only viable option at the moment.  I must emphasize that the gauge field is an integral part of this story, and not a technical artifact of our particular formulation.  In principle it is possible to formulate the problem with spin carrying bosons and fermions and holes, and reach the same low energy state described here.  In this sense this proposed solution of the high T$_c$ problem represents the emergence of new physics.

Two recent developments are encouraging.  First, it appears possible to probe the ``normal state'' at zero temperature by applying a large magnetic field.  While the state again appears to be a Fermi liquid, it may expose a new kind of order and reveal what is fluctuating in the pseudogap phase.  Secondly, the elusive spin liquid ground state may finally have been realized.  Even though this is an early stage in its development, the study of the low energy excitations of these ground states may finally allow us to access spinons and gauge fields.  The effective low energy theory of the spin liquid problem is also more amenable to treatment because bosonic holes are not involved.  It is our view that the study of spin liquids and high $T_c$ opens a new chapter in condensed matter, one that defies the dual Landau paradigm (local order parameter and Fermi liquid theory) that has been the backbone of our Science for three quarters of a century.  To me, this is why the high $T_c$ problem is important and exciting.

\end{enumerate}

\acknowledgements
I thank Naoto Nagaosa and Xiao-Gang Wen for their collaboration on many of the topics discussed here and I am particularly grateful to T. Senthil for sharing his insights and for helping shape many of the thoughts that have gone into this review.  I acknowledge support by NSF grant number DMR--0517222.

\end{document}